\documentclass[twocolumn,english,preprintnumbers,prd,floatfix]{revtex4}

\usepackage{graphicx}
\usepackage{epsfig}
\usepackage{rotate}
\usepackage{latexsym}
\usepackage{amssymb}
\usepackage{psfrag}

\newcommand{\beq}{\begin{equation}}
\newcommand{\eeq}{\end{equation}}
\newcommand{\bdm}{\begin{displaymath}}
\newcommand{\edm}{\end{displaymath}}
\newcommand{\bea}{\begin{eqnarray}}
\newcommand{\eea}{\end{eqnarray}}
\newcommand{\bt}{\begin{tabular}}
\newcommand{\et}{\end{tabular}}
\newcommand{\lan}{\langle}
\newcommand{\ran}{\rangle}
\newcommand{\kv}{{\bf k}}

\newcommand{\qv}{{\bf q}}
\newcommand{\pv}{{\bf p}}
\newcommand{\xv}{{\bf x}}
\newcommand{\Xv}{{\bf X}}
\newcommand{\yv}{{\bf y}}
\newcommand{\zv}{{\bf z}}
\newcommand{\de}{{\rm d}}
\newcommand{\smax}{\mbox{\tiny{max}}}
\newcommand{\smin}{\mbox{\tiny{min}}}
\newcommand{\intkqi}{\int_{k_i}\!\!\!\!\!\de^3 q_1}
\newcommand{\intkqa}{\int_{k_{i_1}}\!\!\!\!\!\!\de^3 q_1}

\newcommand{\intkqc}{\int_{k_{i_3}}\!\!\!\!\!\!\de^3 q_3}
\newcommand{\intkpa}{\int_{k_{j_1}}\!\!\!\!\!\!\de^3 p_1}
\newcommand{\intkpb}{\int_{k_{j_2}}\!\!\!\!\!\!\de^3 p_2}
\newcommand{\intkpc}{\int_{k_{j_3}}\!\!\!\!\!\!\de^3 p_3}

\newcommand{\mnras}{Mon.\ Not.\ R.\ Astron.\ Soc.\ }
\newcommand{\aap}{Astron.\ Astrophys.\ }
\newcommand{\apjs}{Astrophys.\ J.\ Suppl.\  }

\def\d{\delta}
\def\Mpc{\, h^{-1} \, {\rm Mpc}}

\def\kMpc{\, h \, {\rm Mpc}^{-1}}
\def\kmpspMpc{\, {\rm km} \,\, {\rm s}^{-1} \, {\rm Mpc}^{-1}}

\def\fun#1#2{\lower3.6pt\vbox{\baselineskip0pt\lineskip.9pt
        \ialign{$\mathsurround=0pt#1\hfill##\hfil$\crcr#2\crcr\sim\crcr}}}
\def\la{\mathrel{\mathpalette\fun <}}
\def\ga{\mathrel{\mathpalette\fun >}}

\begin{document}

\preprint{FERMILAB-PUB-06-083-A}

\title{Cosmology and the Bispectrum}

\author{Emiliano Sefusatti$^{1,2}$, Mart\'in Crocce$^2$, Sebasti\'an Pueblas$^2$, Rom\'an Scoccimarro$^2$}
\affiliation{${}^{1}$Particle Astrophysics Center, Fermi National Accelerator Laboratory, Batavia, IL 60510-0500}
\affiliation{${}^{2}$Center for Cosmology and Particle Physics, Department of Physics, New York University, New York, NY 10003}

\begin{abstract}

The present spatial distribution of galaxies in the Universe is non-Gaussian, with 40\% skewness in $50 \Mpc$ spheres, and remarkably little is known about the information encoded in it  about cosmological parameters beyond the power spectrum. In this work we present an attempt to bridge this gap by studying the bispectrum, paying particular attention to a joint analysis with the power spectrum and their combination with CMB data. We address the covariance properties of the power spectrum and bispectrum including the effects of beat coupling that lead to interesting cross-correlations, and discuss how baryon acoustic oscillations break degeneracies. We show that the bispectrum has significant information on cosmological parameters well beyond its power in constraining galaxy bias, and when combined with the power spectrum is more complementary than combining power spectra of different samples of galaxies, since non-Gaussianity provides a somewhat different direction in parameter space. In the framework of flat cosmological models we show that most of the improvement of adding bispectrum information corresponds to parameters related to the amplitude and effective spectral index of perturbations, which can be improved by almost a factor of two. Moreover, we demonstrate that the expected statistical uncertainties in $\sigma_8$ of a few percent are robust to relaxing the dark energy beyond a cosmological constant. 

\end{abstract}

\maketitle

\section{Introduction}

Several recent studies have stressed the importance of combining different observations to constrain cosmological parameters. A clear example is provided by the analysis of the galaxy power spectrum in the Sloan Digital Sky Survey (SDSS)~\cite{Tegmark:2003ud}, and in the 2dF Galaxy Redshift Survey~\cite{Sanchez:2005pi}, which have shown the central role played by the information contained in the large-scale galaxy distribution to break the degeneracies still present in the cosmic microwave background (CMB) data despite the precision of the WMAP satellite observations~\cite{Spergel:2003cb,Spergel:2006hy}. 

One of the main challenges in extracting cosmological information from galaxy clustering is knowing how good tracers of the underlying mass distribution galaxies are. This is often bypassed altogether, for example in~\cite{Tegmark:2003ud,Sanchez:2005pi} only infomation on the {\it shape} of the galaxy power spectrum was used, since its amplitude is degenerate with the linear bias parameter relating galaxy to dark matter fluctuations at large scales. 

The determination of galaxy bias has been, so far, among the main reasons of interest in the galaxy higher-order statistics in general~\cite{Frieman:1993nc,Frieman:1999qj,Szapudi02,Gazta02,TaJa03,Croton04,JiBo04,Kayo04,Wang04,Pan:2005ym,Gazta05,FPS05} and the bispectrum in particular~\cite{Fry1994,Scoccimarro:2000sp,Feldman:2000vk,Verde:2002ed,SWS05}. At large scales, the dependence on triangle configuration of the bispectrum generated by gravitational instability  allows to disentagle the gravitational contribution from the bispectrum generated by non-linear biasing and ultimately remove the degeneracy between the linear bias and the amplitude of the dark matter perturbations. In weak gravitational lensing at smaller scales, the bispectrum can similarly be used to break degeneracies between matter content and the amplitude of fluctuations and probe dark energy~\cite{BWM97, Hui99, CoHu01, TaJa04, Kilbinger:2005jy}.

Observational applications of this method to the galaxy distribution in the past involved fixing the cosmological model~\cite{Scoccimarro:2000sp,Feldman:2000vk,Verde:2002ed,Gazta05}, thus the information of the bispectrum on cosmological parameters has not been properly taken advantage of. In this work we study the constraining power of the bispectrum as a tool in the determination of cosmological parameters and the nature of primordial fluctuations, going beyond the determination of galaxy bias alone. As shown in~\cite{Sefusatti:2004xz}, higher-order correlation functions such as the bispectrum or the trispectrum in galaxy surveys show, when all measurable configurations are taken into account, a signal-to-noise ratio comparable or even exceeding the signal-to-noise of the power spectrum at mildly non-linear scales.

\begin{figure}[t]
\begin{center}
\includegraphics[width=0.48\textwidth]{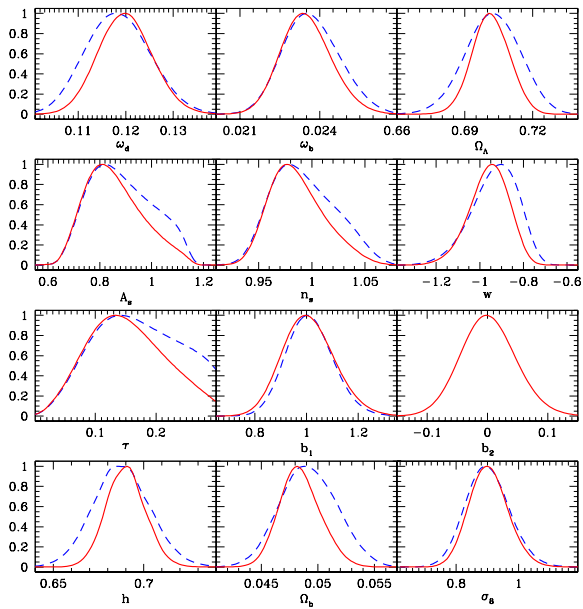}
\caption{\label{figPvsB} An example showing the constraining power of the bispectrum compared to the power spectrum. The panels show marginalized likelihood functions corresponding to a hypothetical joint analysis of WMAP (first year) and SDSS North (by the end of the survey) where only the galaxy power spectrum is used (blue, dashed line), or only the galaxy bispectrum is used (red, solid line). Assumes a flat cosmology and scales up to $k_{\smax}=0.3\kMpc$.}
\end{center}
\end{figure}

Postponing a detailed discussion of our method to the following sections, we show in Fig.~\ref{figPvsB} how the power spectrum and the bispectrum measured from the same data set compare in constraining cosmological parameters. We consider flat cosmological models depending on nine parameters: the physical dark matter density $\omega_d=\Omega_d h^2$, the physical baryon density $\omega_b=\Omega_b h^2$, the dark energy density $\Omega_\Lambda$, the amplitude of scalar fluctuations $A_s$, the scalar spectral index $n_s$, the dark energy equation of state parameter $w$, the optical depth to Thomson scattering $\tau$,  plus the linear ($b_1$) and quadratic ($b_2$) galaxy bias parameters. We also show ``derived" parameters such as $h$, the Hubble constant in units of $100\kmpspMpc$, the baryon density $\Omega_b$ and the amplitude of dark matter fluctuations at $8\Mpc$, $\sigma_8$. 

These constraints are from a hypothetical analysis that combines the CMB data from WMAP (first year) with measurements in the North part of SDSS by the end of the survey in two cases: using the SDSS galaxy power spectrum (blue, dashed line) and {\em replacing} the SDSS power spectrum by the bispectrum (red, continuous line). Both cases include the covariance between different power spectrum bins or bispectrum configurations (see below for a full discussion). Figure~\ref{figPvsB} shows that when all triangle configurations are included down to wavenumber $k_{\smax}=0.3\kMpc$ the bispectrum even improves the power spectrum results. 

In practice one would like to combine the information in the power spectrum and bispectrum, which requires a calculation of their covariance properties. This is the main subject of the present work.  The cross-covariance between the power spectrum and bispectrum turns out to have some non-trivial properties that help constraining cosmological parameters, in a way that is unexpected from a separate consideration of the covariance of each statistic by itself.

Although the main properties of the covariance matrices can be understood analytically, the details of the survey under consideration are important in practice, thus we compute covariance matrices from mock catalogs designed to reproduce the geometry of the SDSS survey by its completion. In particular, we consider the power spectrum and bispectrum of the north hemisphere main sample (MS) of galaxies. We also discuss how our constrains change as we include the power spectrum of the Luminous Red Galaxies (LRG) sample in the same geometry. As an example of what should be expected in combining large-scale structure (LSS) with the CMB we use the WMAP first year data; after the present work appeared as a preprint the WMAP 3-year  (WMAP3) data became available~\cite{Spergel:2006hy}. The new data provides a somewhat different angle on the constraining power of the bispectrum, not just incremental improvements, for this reason we consider separately in the Appendix what happens when WMAP3 is added.

This paper is organized as follows. In section~\ref{secBisp} we review some basic results regarding the large-scale bispectrum of the galaxy distribution and discuss the main features of the covariance measured from our mock catalogues in section~\ref{COVA}. In section~\ref{secLikelihood} we present the likelihood functions both for the LSS and CMB correlations. In section~\ref{secResults} we present results on expected constraints on cosmological parameters, derived assuming the WMAP 1-year data (WMAP1). We conclude in section~\ref{secConclusions}. The Appendix discusses what happens when we replace WMAP1 by WMAP3.

\section{Predictions and Mock Catalogs}
\label{secBisp}

We will assume primordial fluctuations to be Gaussian, so that every connected higher-order correlation function in the dark matter overdensity field $\d$ results from gravitational instability. The dark matter bispectrum $B$, {\em i.e.} the Fourier counterpart of the 3-point correlation function, is defined as the ensemble average
\beq
\lan \d_{\kv_1}\d_{\kv_2}\d_{\kv_3} \ran\equiv\d_D(\kv_{123})\ B(k_1,k_2,k_3),
\label{BispTree}
\eeq
with $\d_\kv$ the density contrast in Fourier space and $\kv_{123}\equiv\kv_1+\kv_2+\kv_3$. If the bispectrum can be reliably predicted by tree-level perturbation theory (PT), it follows that~\cite{Fry84}
\beq
B(k_1,k_2,k_3)=2 F_2(\kv_1,\kv_2)P_1 P_2+{\rm cyc.},
\label{BispEPT}\eeq
where $P_1\equiv P(k_1)$ is the linear power spectrum while the kernel $F_2$ reads
\beq
F_2(\kv_1,\kv_2)=\frac{5}{7}+\frac{x}{2}\left(\frac{k_1}{k_2}+\frac{k_2}{k_1}\right)+\frac{2}{7}x^2,
\eeq
with $x=\hat{\kv}_1\cdot\hat{\kv}_2$. 

A second source of non-Gaussianity in the galaxy density field is given by non-linear galaxy bias. At scales much larger than the typical size of virialized structures the relation between the galaxy distribution and the underlying dark matter distribution is  expected to be local~\cite{Fry:1992vr,Coles93,ScWe98}, that is, in terms of the respective overdensities, $\delta_g(x)=f[\delta(x)]$. For small fluctuations we can Taylor-expand and describe such function in terms of few, constant, bias parameters~\cite{Fry:1992vr}
\beq
\delta_g(x)= b_1\delta+\frac{1}{2}b_2\delta^2+\frac{1}{3!}b_3\delta^3+...
\label{locbias}
\eeq 
The large-scale galaxy power spectrum will therefore be given $P_g(k)\simeq b_1^2 P(k)$ while the galaxy bispectrum $B_g$ will be related to dark matter bispectrum $B$ as
\beq\label{Bgal}
B_g(k_1,k_2,k_3)\simeq b_1^3B(k_1,k_2,k_3)+b_1^2 b_2 (P_1P_2+{\rm cyc.})
\eeq
The different behaviour of the first and second terms on the right hand side of Eq.~(\ref{Bgal}) as a function of triangle configuration given by the wavenumbers $k_1$, $k_2$ and $k_3$ allows a simultaneous measurement of the linear bias parameter $b_1$ and the quadratic bias parameter $b_2$~\cite{Frieman:1993nc,Fry1994}. This becomes obvious when Eq.~(\ref{Bgal}) is rewritten in terms of the reduced bispectrum, defined, for the dark matter field as $Q\equiv B(k_1,k_2,k_3)/(P_1 P_2+ {\rm cyc.})$ and analogously for the galaxy distribution, so that
\beq\label{Qgal}
Q_g(k_1,k_2,k_3)=\frac{1}{b_1}Q(k_1,k_2,k_3)+\frac{b_2}{b_1^2}.
\eeq
While the first term on the left hand side depends on the specific triangle via the $F_2(\kv_1,\kv_2)$ kernel, the second just amounts to an overall additive constant. 

As mentioned above, the scale-dependence of the bias parameter is expected to be very weak at the scales considered in the present analysis. This can be shown in the framework of the halo model and it can probed, observationally, by looking at the dependence of measured values of $b_1$ and $b_2$ on the smallest scale, or largest wavenumber $k_{\smax}$, included in the analyis. If there is a scatter about the deterministic relationship given by Eq.~(\ref{locbias}), the bispectrum method recovers the {\em mean} relationship between galaxy and matter overdensities. This has been shown for models with significant scatter (see Fig.~1 in~\cite{Sco00b}) and galaxies populated using Halo Occupation Distributions (HOD) where the scatter is typically not very significant at the scales we consider here (see Fig.~6 in~\cite{GaSc05}).

In this work we consider scales up to $k \leq 0.3\kMpc$, for which the validity of  Eq.~(\ref{BispEPT}) is only accurate to about 20$\%$~\cite{Scoccimarro:1997st,Sco00b}. A more accurate description of the bispectrum at these scales, particularly in redshift space, is given by second-order Lagrangian PT (2LPT), \cite{Sco00b}. Therefore, we will only use tree-level PT to model {\em deviations} from a fiducial model calculated by using mock catalogs generated by {\tt PTHalos}~\cite{SS02} and 2LPT simulations, which are similar at these scales, since halos in {\tt PTHalos} are placed in the large-scale 2LPT density field. The advantage of using {\tt PTHalos} is that a biased population of galaxies can be chosen by using appropriate prescriptions for their occupation inside halos. This method is therefore necessary for LRG galaxies, which are strongly biased tracers, whereas main sample galaxies are close enough to being unbiased that the difference between using 2LPT and {\tt PTHalos} is not significant. 

We therefore use the mock catalogs for the main sample of galaxies in SDSS generated by using 2LPT in~\cite{Scoccimarro:2003wn}, where the following cosmological parameters were used: dark matter density $\Omega_d=0.225$, baryon density $\Omega_b=0.045$, cosmological constant with density $\Omega_\Lambda=0.73$, Hubble constant $h=0.71$ and fluctuation amplitude $\sigma_8=0.82$ at the mean redshift of the survey of $z_{\rm mean} \simeq 0.1$. As discussed above, we assume these galaxies to be unbiased, and have included the detailed geometry of the expected final angular and radial selection functions. The redshift-space density field is weighted using the Feldman-Kaiser-Peacock (FKP) procedure~\cite{Feldman:1993ky,Sco00b,MVH97} with $P_0=5000~(\Mpc)^3$. We measure the power spectrum from $k_{\smin}=0.02\kMpc$ to $k_{\smax}=0.31\kMpc$, with a bin size given by $\Delta k=0.015\kMpc$. We consider $N_k=20$ $k$-bins for the power spectrum, corresponding to $N_T=1015$ triangle bins, including all triangle shapes and orientations corresponding to $7.5 \times 10^{10}$ elementary triangles. We use  $6000$ realizations of the survey~\cite{Scoccimarro:2003wn,Sefusatti:2004xz}, such a large number is needed in order to estimate covariance matrices larger than $10^3\times10^3$ in size (see next section). 

For the mock catalogs of the LRG sample, we use $6000$ mock catalogs constructed with  {\tt PTHalos}, using the following cosmological parameters: dark matter density $\Omega_d=0.229$, baryon density $\Omega_b=0.046$, cosmological constant with density $\Omega_\Lambda=0.725$, Hubble constant $h=0.71$ and fluctuation amplitude $\sigma_8=0.75$ at the mean redshift of the survey of $z_{\rm mean} \simeq 0.35$. In these mock catalogs the LRG galaxies populate dark matter halos according to an HOD prescription~\cite{BeWe02} for the mean number of galaxies in a halo of mass $m$
\beq
\langle N_{\rm gal}(m)\rangle =  {\rm e}^{-m_{\rm min}/m}\ \Big[1+\Big(\frac{m}{m_1}\Big)^\alpha \Big],
\label{HOD}
\eeq
where the first contribution is that due to a central galaxy (with nearest integer scatter), the rest being satellite galaxies which are taken with a Poisson distributed scatter~\cite{Kravtsov04}. The parameters are chosen by a best fit procedure of the large-scale redshift-space correlation function given in~\cite{Eisenstein:2005su} (including the survey covariance matrix) and the small-scale redshift-space correlation function given in~\cite{Zehavi2005}. The resulting parameters are $m_{\rm min}=5 \times 10^{13}\,M_\odot h^{-1}$, $m_1= 10^{15}\, M_\odot h^{-1}$ and $\alpha=1.95$. 
Given Eq.~(\ref{HOD}), the large-scale bias parameters are given by,
\beq
b_i = \frac{1}{n_g} \int dm\ n(m)\ \langle N_{\rm gal}(m) \rangle\ b_i(m),
\label{bhod}
\eeq
where $n(m)$ is the halo mass function (assumed to be that in~\cite{ShTo02}), $b_i(m)$ are the corresponding halo bias parameters~\cite{SSHJ01} and the galaxy number density is given by
\beq
n_g = \int dm\ n(m)\ \langle N_{\rm gal}(m) \rangle .
\label{ngal}
\eeq
For the parameters given above, $n_g=8\times 10^{-5}$, $b_1=2.11$, $b_2= 1.1$ and $b_3= -2.8$. In practice, the values of the bias parameters measured in the mock catalogs are slightly different from the analytical calculations due to the particular prescription adopted to describe how individual halos are populated. We find, for example, $b_1=2.17$ and  we use this as our fiducial value for the LRG large-scale linear bias. The mock catalogs have the radial selection function expected by the end of the SDSS survey and described in \cite{Eisenstein:2005su} and an angular selection function equal to unity inside the survey region. The redshift-space density field in them is weighted according to the FKP method with $P_0=40,000~(\Mpc)^3$.

In section~\ref{secLikelihood} below we discuss in more detail how we take into account redshift distortions. In brief, we assume the 2LPT or {\tt PTHalos} simulations give the correct answer (note that these {\em do not} assume perturbation theory for the real-to-redshift space mapping), and compute {\em deviations} from the fiducial model by tree-level perturbation theory. A full discussion about accurate theoretical predictions for statistics of galaxies in redshift-space and their possible systematics is beyond the scope of this paper, and will be presented elsewhere.

\section{Covariance matrices}
\label{COVA}

In order to perform a joint likelihood analysis of the power spectrum and bispectrum, as detailed in section~\ref{secLikelihood} below, we need to compute their covariance properties. The {\em full} covariance matrix $C_{ij}$ obtained  from our mocks catalogs by measuring the power spectrum and bispectrum, is defined as
\beq
C_{ij}\equiv \lan \delta X_i \delta X_j \ran
\eeq
where $\delta X_i=X_i-\bar{X}_i$ and $X_i$ equals the power spectrum $P_i$ for $i=1,...,N_k$ with $N_k$ the number of power spectrum bins, or the bispectrum for $i=N_k+1,...,N_k+N_T$, with $N_T$ the number of bins in triangle space. 

In what follows, we consider the three contributions to the general covariance matrix $C_{ij}$, that is $\lan\d P_i \d P_j\ran$, $\lan\d B_i \d B_j\ran$ and $\lan\d P_i \d B_j\ran$ in turn, and compare the expected contributions to the values measured from the mock catalogues.

\subsection{Power Spectrum Covariance}
\label{secPScov}

\begin{figure*}[th!]
\begin{center}
\bt{cc}
\includegraphics[width=0.49\textwidth]{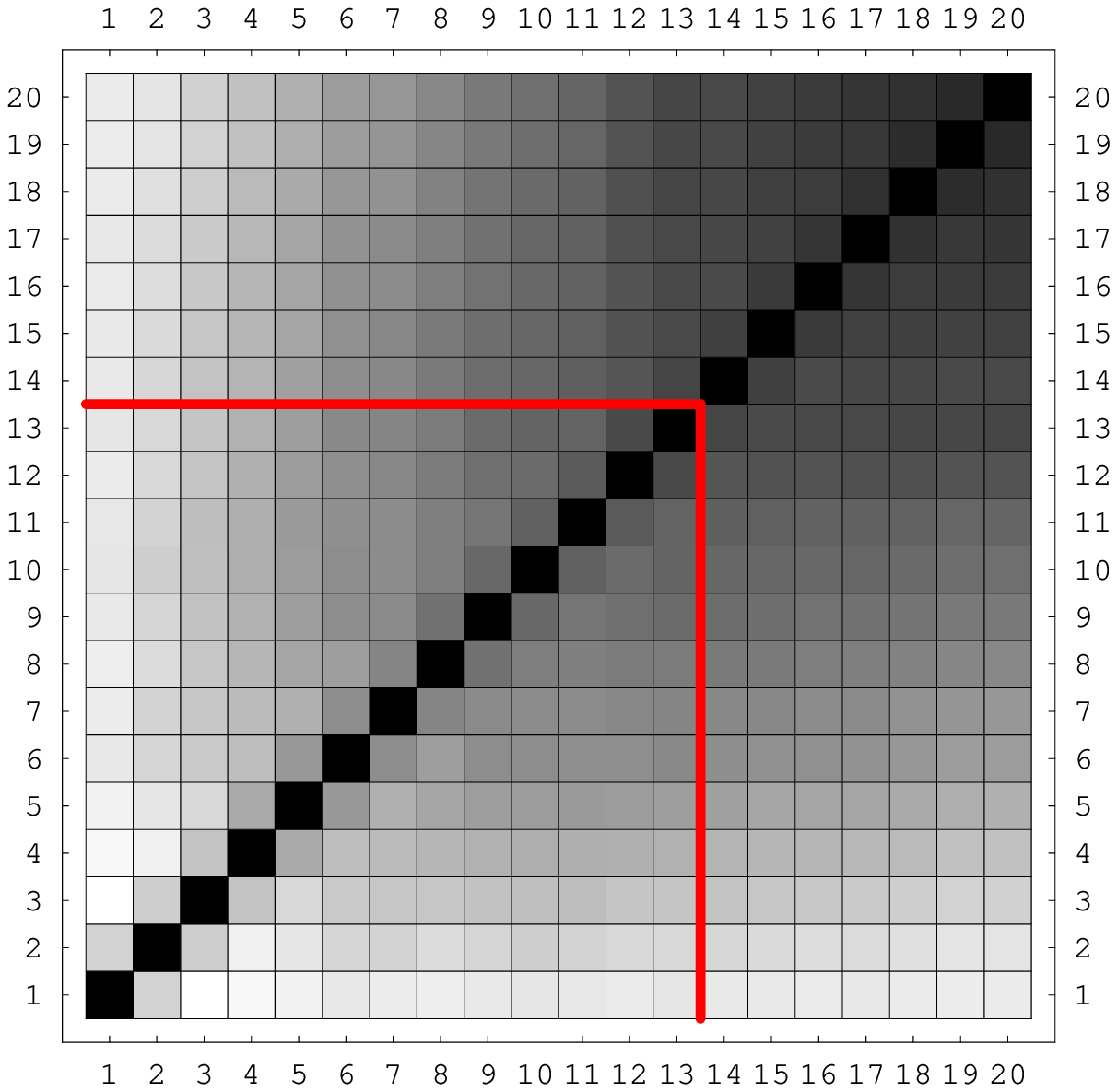}&
\includegraphics[width=0.49\textwidth]{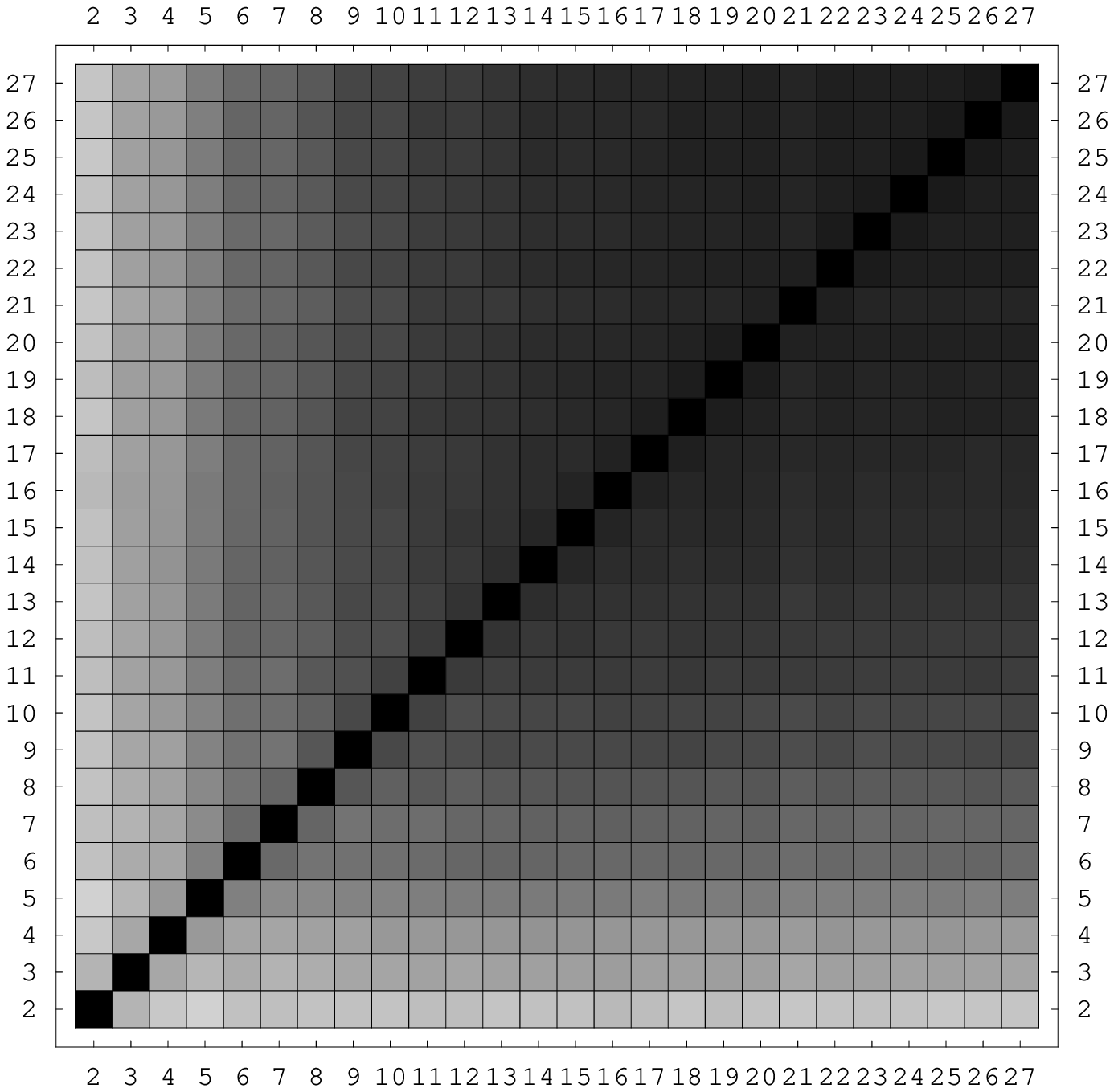}\\
\et
\caption{\label{densityPS} Power spectrum cross-correlation coefficients $r_{ij}^P$ between different scales for the main (left) and LRG sample (right) as measured from the SDSS mock catalogs. Black indicates maximum cross-correlation ($r_{ij}^P=1$), white no cross-correlation ($r_{ij}^P=0$). The wavenumbers are given in units of $\Delta k\simeq 0.015\kMpc$ for the main sample power spectrum, while $\Delta k\simeq 0.0075\kMpc$ for the LRG power spectrum. Note that bin 27 in the LRG case corresponds to bin 13 in the main sample case: the lower left region in the main sample plot encloses the scales corresponding to the LRG plot.}
\end{center}
\end{figure*}
\begin{table}[b!]
\caption{\label{table_ccPS} Power spectrum cross-correlation coefficients $C_{ij}^P$ between different scales as measured from the main sample SDSS mock catalogues. The value of the wavenumber $k$ for the corresponding bin is given in the first column in units of $\kMpc$. For brevity, only even bins are displayed.}
\begin{ruledtabular}
\begin{tabular}{ll|llllllllll}
$k$     & bin &$2$     & $4$    &$6$     & $8$    &$10$    & $12$    & $14$    & $16$    & $18$    &$20$ \\     
\hline
$0.031$ & 2   &$1.00$  &$0.13$  &$0.22$  &$0.2$   &$0.25$  &$0.21$  & $0.22$ & $0.2$  & $0.18$  & $0.17$ \\
$0.063$ & 4   &$      $&$1.00$  &$0.31$  &$0.34$  &$0.37$  &$0.36$  & $0.34$ & $0.34$ & $0.32$  & $0.30$ \\
$0.094$ & 6   &$      $&$      $&$1.00$  &$0.42$  &$0.49$  &$0.48$  & $0.48$ & $0.47$ & $0.45$  & $0.43$ \\
$ 0.126$& 8   &$      $&$      $&$      $&$1.00$  &$0.54$  &$0.55$  & $0.55$ & $0.54$ & $0.53$  & $0.51$ \\
$0.157$ & 10  &$      $&$      $&$      $&$      $&$1.00$  &$0.61$  & $0.62$ & $0.62$ & $0.61$  & $0.60$ \\
$0.188$ & 12  &$      $&$      $&$      $&$      $&$      $&$1.00$  & $0.69$ & $0.7$  & $0.71$  & $0.70$ \\
$0.220$ & 14  &$      $&$      $&$      $&$      $&$      $&$      $& $1.00$ & $0.73$ & $0.74$  & $0.74$ \\
$0.251$ & 16  &$      $&$      $&$      $&$      $&$      $&$      $&$      $& $1.00$ & $0.78$  & $0.78$ \\
$0.283$ & 18  &$      $&$      $&$      $&$      $&$      $&$      $&$      $&$      $& $1.00$  & $0.82$ \\
$0.314$ & 20  &$      $&$      $&$      $&$      $&$      $&$      $&$      $&$      $& $     $ & $1.00$ \\
\end{tabular}
\end{ruledtabular}
\end{table}
Our power spectrum estimator can be written as
\beq\label{PSest}
\hat{P}(k)\equiv\frac{k_f^3}{V_P(k)}\int_{k}\!\!\de^3 q_1\int_{k}\!\!\de^3 q_2\;\d_D(\qv_{12})\;\d_{\qv_1}\d_{\qv_2},
\eeq
where 
\beq
V_P(k)\equiv\int_{k}\!\!\de^3 q_1\int_{k}\!\!\de^3 q_2\;\delta_D(\qv_{12})\simeq 4\pi k^2\Delta k,
\eeq
and the integral over the bin $k$ of width $\Delta k$ is given by
\beq
\int_k\!\!\de^3 q\equiv\int_{k-\Delta k/2}^{k+\Delta k/2}\!\!\de k\, k^2\int\de \Omega.
\eeq
\begin{table}[b!]
%\caption{\label{table_ccPS_LRG} Power spectrum cross-correlation coefficients between different scales as measured from the LRG sample SDSS mock catalogues. The value of the wavenumber $k$ for the corresponding bin is given in the first column in units of $\kMpc$. For brevity, only one bin every three is shown.}
\caption{\label{table_ccPS_LRG} Same as Table~\ref{table_ccPS} but for the LRG sample SDSS mock catalogues. For brevity, only one bin every three is shown.}
\begin{ruledtabular}
\begin{tabular}{ll|lllllllll}
$k$     & bin & 2      & 5      & 8      & 11      & 14     & 17     & 20     & 23     & 26 \\
\hline
$0.015$ & 2  & $1.00$   & $0.18$   & $0.24$   & $0.26$   & $0.24$   & $0.26$   & $0.24$   & $0.25$   & $0.23$ \\
$0.038$ & 5  & $    $   & $1.00$   & $0.46$   & $0.51$   & $0.52$   & $0.51$   & $0.52$   & $0.5$   & $0.5$ \\
$0.060$ & 8  & $    $   & $   $   & $1.00$   & $0.65$   & $0.66$   & $0.66$   & $0.66$   & $0.64$   & $0.66$ \\
$0.083$ & 11 & $    $   & $   $   & $  $   & $1.00$   & $0.77$   & $0.77$   & $0.77$   & $0.76$   & $0.78$ \\
$0.106$ & 14 & $    $   & $   $   & $$   & $  $   & $1.00$   & $0.83$   & $0.83$   & $0.82$   & $0.83$ \\
$0.123$ & 17 & $    $   & $   $   & $ $   & $ $   & $ $   & $1.00$   & $0.85$   & $0.84$   & $0.84$ \\
$0.151$ & 20 & $    $   & $   $   & $ $   & $ $   & $ $   & $ $   & $1.00$   & $0.86$   & $0.87$ \\
$0.173$ & 23 & $    $   & $   $   & $ $   & $ $   & $ $   & $ $   & $ $   & $1.00$   & $0.87$ \\
$0.196$ & 26 & $    $   & $   $   & $ $   & $ $   & $ $   & $ $   & $ $   & $ $   & $1.00$ \\
\end{tabular}
\end{ruledtabular}
\end{table}
The bin width $\Delta k$ does not necessarily coincide with the fundamental frequency $k_f$ (in our analysis we will consider the case $\Delta k=3 k_f$). If the value of the power spectrum averaged over all the realizations is $P(k)=\lan\hat{P}(k)\ran$, it is easy to see that the covariance between power spectrum bins can be expressed as~\cite{SZH99, MeWh99}
\bea
C_{ij}^P & \equiv & \lan \d P(k_i) \d P (k_j) \ran =\nonumber\\
 & \simeq & \d_{ij}\ \frac{2~k_f^3}{V_P(k_i)}P^2(k_i)\nonumber\\
 & + & \frac{k_f^3}{2}\int_{-1}^{+1}\!\!\!\!\!\de\cos\theta\ \widetilde{T}(k_i,k_j,\theta)
 \label{Pcova}
\eea
where the first diagonal term is the Gaussian contribution and in the second, non-Gaussian, term $\widetilde{T}(k_i,k_j,\theta)=T(\kv_i,-\kv_i,\kv_j,-\kv_j)$ is the trispectrum of the dark matter field and $\theta$ is the angle between the vectors $\kv_i$ and $\kv_j$. 

Note that expressions such as Eq.~(\ref{Pcova}), or the analogue ones for the bispectrum and mixed covariance discussed in the next sections, assume that the survey window is effectively a delta function in Fourier space, and they are therefore exact in the case of a periodic box. However, they provide a simple estimate of the covariance for more generic survey geometries {\em except} for the case of the mixed power spectrum - bispectrum covariance, $\lan \delta P_i \delta B_j\ran$ as we will see in section~\ref{secPBcov} below. Of course, in analyzing our mock catalogs we do not make any such approximation as the survey geometry is properly taken into account by the FKP method. The estimators for power spectrum and bispectrum will therefore include convolutions with the window function and shot noise terms and the covariance will then be computed from the measured statistics in each mock catalogue.

Figure~\ref{densityPS} shows the redshift-space power spectrum cross-correlation coefficients
\beq
r_{ij}^P\equiv\frac{C_{ij}^P}{\sqrt{C_{ii}^P C_{jj}^P}},
\label{rijP}
\eeq
for the main (left) and LRG (right) sample power spectra measured from our mock catalogs. The values are ranging from $0$ (white) to $1$ (black). We used $N_k=20$ bins for the main sample, $N_k^{LRG}=27$ bins in the LRG sample case. The numerical value of the cross-correlation coefficients for the main sample power spectrum is given in Table~\ref{table_ccPS}, whereas Table~\ref{table_ccPS_LRG} presents the LRG power spectrum case. Note that in this case the maximum value for the wavenumber considered is $k_{\smax}^{LRG}=0.2\kMpc$, instead of $k_{\smax}^{MS}=0.3\kMpc$ for the main sample.

As evident from these tables and Fig.~\ref{densityPS}, the cross-correlation between different scales is stronger for the LRG power spectrum case. For example let's consider for the main sample the coefficient $C_{4,8}^P=0.34$ (Tab.~\ref{table_ccPS}) corresponding to the wavenumbers $0.126$ and $0.63\kMpc$ and compare it to the LRG coefficient $C_{8,17}^P=0.66$ (Tab.~\ref{table_ccPS_LRG}) corresponding to the wavenumbers $0.123$ and $0.60\kMpc$. If LRG galaxies were simply a linearly biased population compared to the main galaxy sample (here assumed to be unbiased), then one would expect exactly the {\em opposite} given our choice of bin widths, that is a smaller value in the LRG case. This is so because the cross-correlation coefficient is independent of the volume of the sample (which appears in Eq.~(\ref{Pcova}) only through $k_f$), proportional to the amount of non-Gaussianity (here given by the averaged trispectrum divided by the power spectrum squared), and proportional to the bin width $\Delta k$. The reason for this last dependence is that the non-Gaussian noise term does not get beaten down by bin averaging, whereas the Gaussian one (which dominates in the denominator in Eq.~(\ref{rijP})) does~\cite{SZH99}. Since our bin size for LRG power is half that of the main sample, and linear bias does not alter non-Gaussianity (and redshift-distortions do so very slightly at large scales, and furthermore $\sigma_8$ is somewhat larger in the main sample), this would give that LRG power should be cross-corrrelated about half as much as main sample galaxy power.

\begin{table}[b!]
\caption{\label{table_ccBS} Bispectrum cross-correlation coefficients for triangles at the largest scales. Each triangular configuration is given in terms of the three vectors $k_1$, $k_2$, $k_3$ in units of the $k$-bin width, $\Delta k\simeq 0.015 \kMpc$.}
\begin{ruledtabular}
\begin{tabular}{l|llllllllll}
triangle & 1 & 2 & 2     & 2    & 3    & 3    & 3    & 3    & 3    & 3 \\  
         & 1 & 1 & 2     & 2    & 1    & 2    & 2    & 3    & 3    & 3 \\  
         & 1 & 1 & 1     & 2    & 1    & 1    & 2    & 1    & 2    & 3 \\  
\hline
1,1,1 & $1.00$ & $0.51$   & $0.33$   & $0.06$   & $0.07$   & $0.28$   & $0.06$   & $0.35$ & $0.06$   & $0.00$ \\
2,1,1 & $    $ & $1.00$   & $0.58$   & $0.11$   & $0.28$   & $0.39$   & $0.14$   & $0.35$ & $0.09$   & $0.00$ \\
2,2,1 & $    $ & $    $   & $1.00$   & $0.32$   & $0.15$   & $0.52$   & $0.27$   & $0.4$  & $0.19$   & $0.03$ \\
2,2,2 & $    $ & $    $   & $    $   & $1.00$   & $0.04$   & $0.17$   & $0.41$   & $0.04$ & $0.22$   & $0.03$ \\
3,1,1 & $    $ & $    $   & $    $   & $    $   & $1.00$   & $0.37$   & $0.11$   & $0.09$ & $0.07$   & $0.04$ \\
3,2,1 & $    $ & $    $   & $    $   & $    $   & $    $   & $1.00$   & $0.41$   & $0.54$ & $0.23$   & $0.07$ \\
3,2,2 & $    $ & $    $   & $    $   & $    $   & $    $   & $    $   & $1.00$   & $0.15$ & $0.44$   & $0.12$ \\
3,3,1 & $    $ & $    $   & $   $    & $    $   & $    $   & $    $   & $    $   & $1.00$ & $0.26$   & $0.09$ \\
3,3,2 & $    $ & $    $   & $    $   & $    $   & $    $   & $    $   & $    $   & $   $  & $1.00$   & $0.34$ \\
3,3,3 & $ $    & $ $      & $    $   & $    $   & $    $   & $    $   & $    $   & $   $  & $    $   & $1.00$ \\ 
\end{tabular}
\end{ruledtabular}
\end{table}

\begin{figure*}[t]
\begin{center}
\bt{cc}
\includegraphics[width=0.49\textwidth]{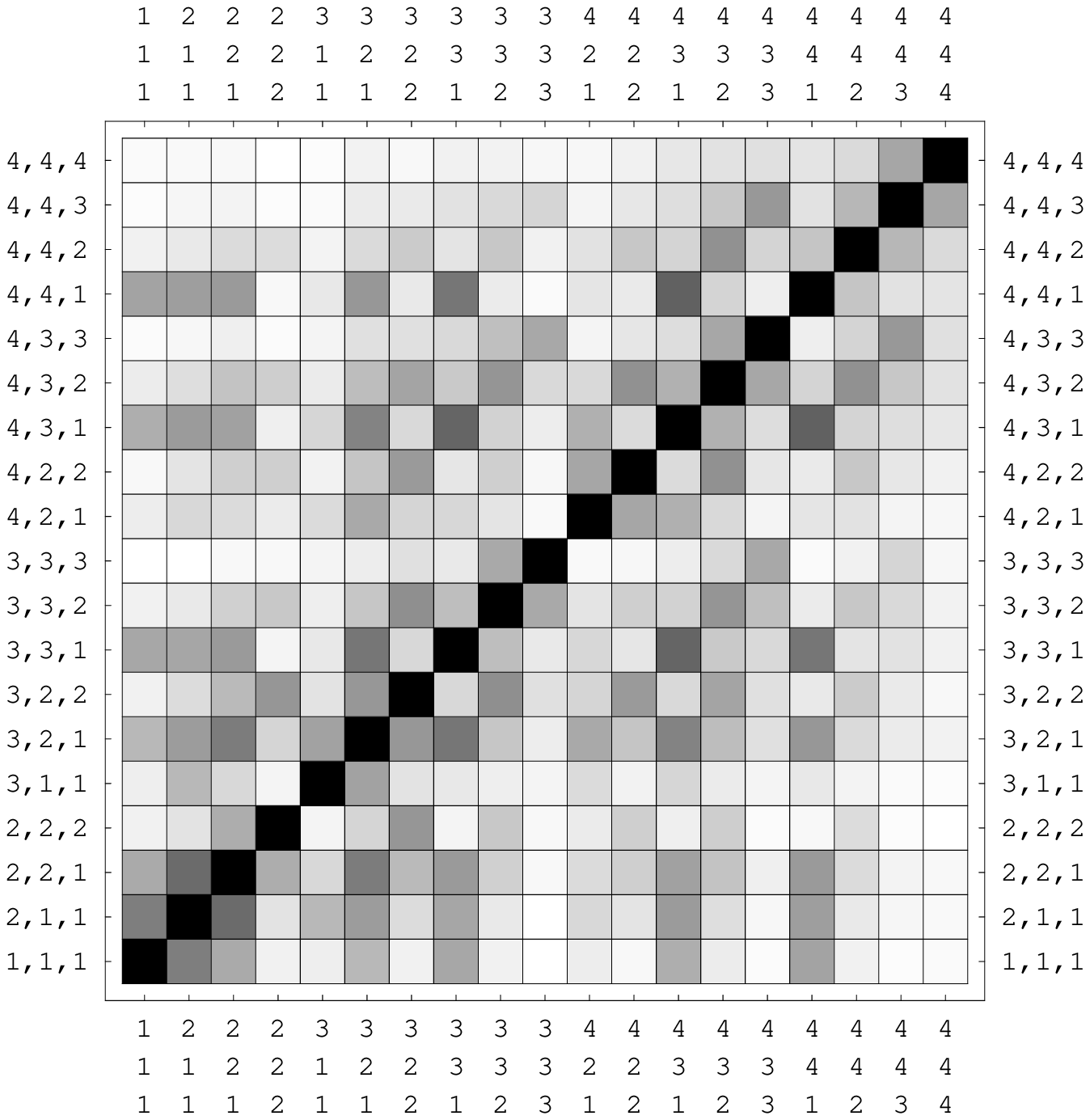}&
\includegraphics[width=0.49\textwidth]{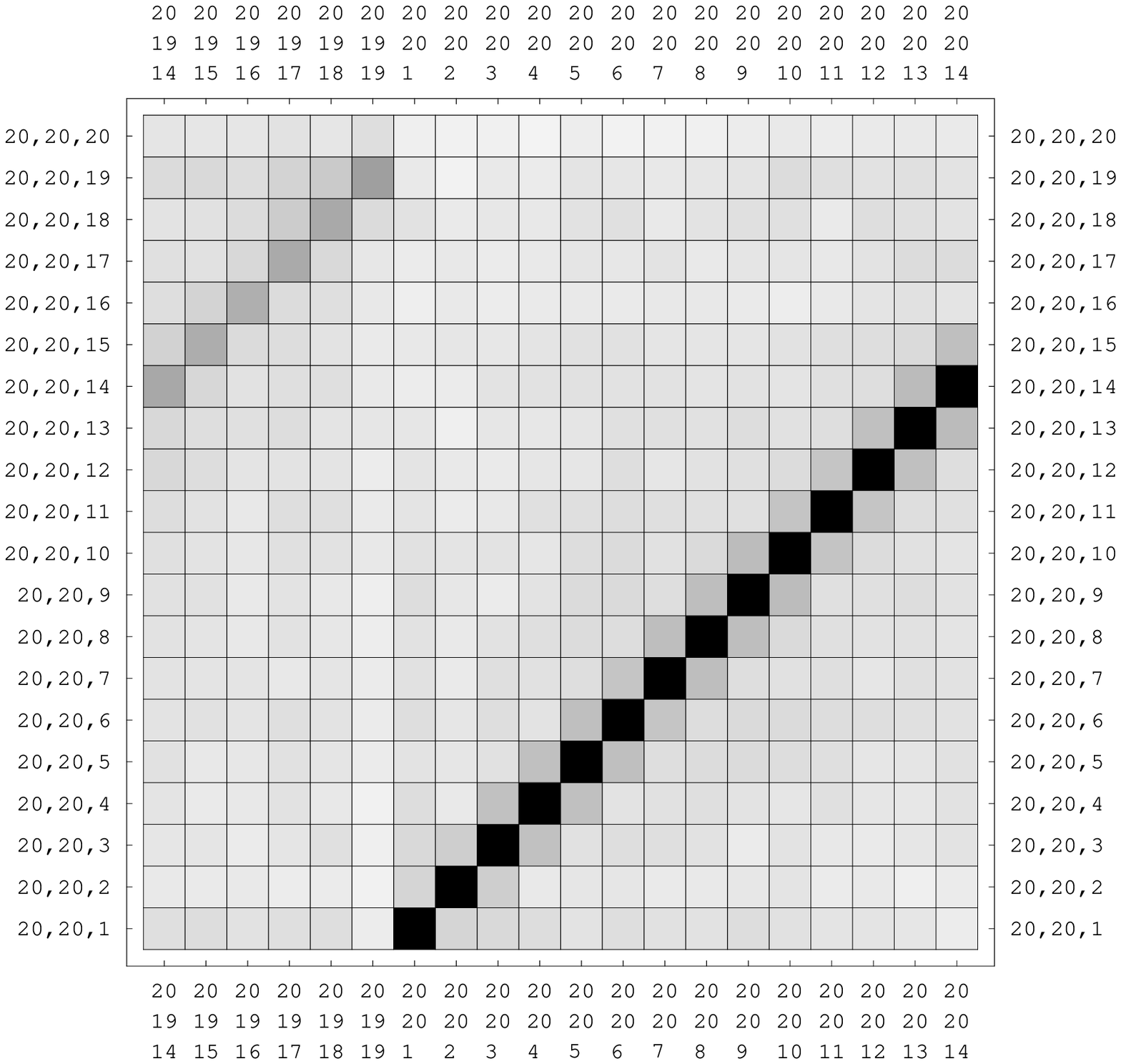}\\
\et
\caption{\label{densityB}Bispectrum cross-correlation coefficients $r_{ij}^B$ among the first $19$ triangles at large scales (left), and among two sets of the $20$ triangles at the smallest scales we consider (right). The triplets indicate the wavenumbers of the triangles sides in units of the $k$-bin width, $\Delta k\simeq0.015\kMpc$.}
%\bt{cc}
%\includegraphics[width=0.49\textwidth]{fig_ccBS_1_20_1_20.eps}&
%\includegraphics[width=0.49\textwidth]{fig_ccBS_995_1015_995_1015.eps}\\
%\et
%\caption{\label{densityB}Bispectrum cross-correlation coefficients $r_{ij}^B$ among the first $20$ triangle at large scales (left), and among the last $25$ triangles at the smallest scales we consider (right). The triplets indicate the wavenumbers of the triangles sides in units of the $k$-bin width, $\Delta k\simeq0.015\kMpc$.}
\end{center}
\end{figure*}
\begin{figure*}[th]
\begin{center}
\bt{cc}
\includegraphics[width=0.49\textwidth]{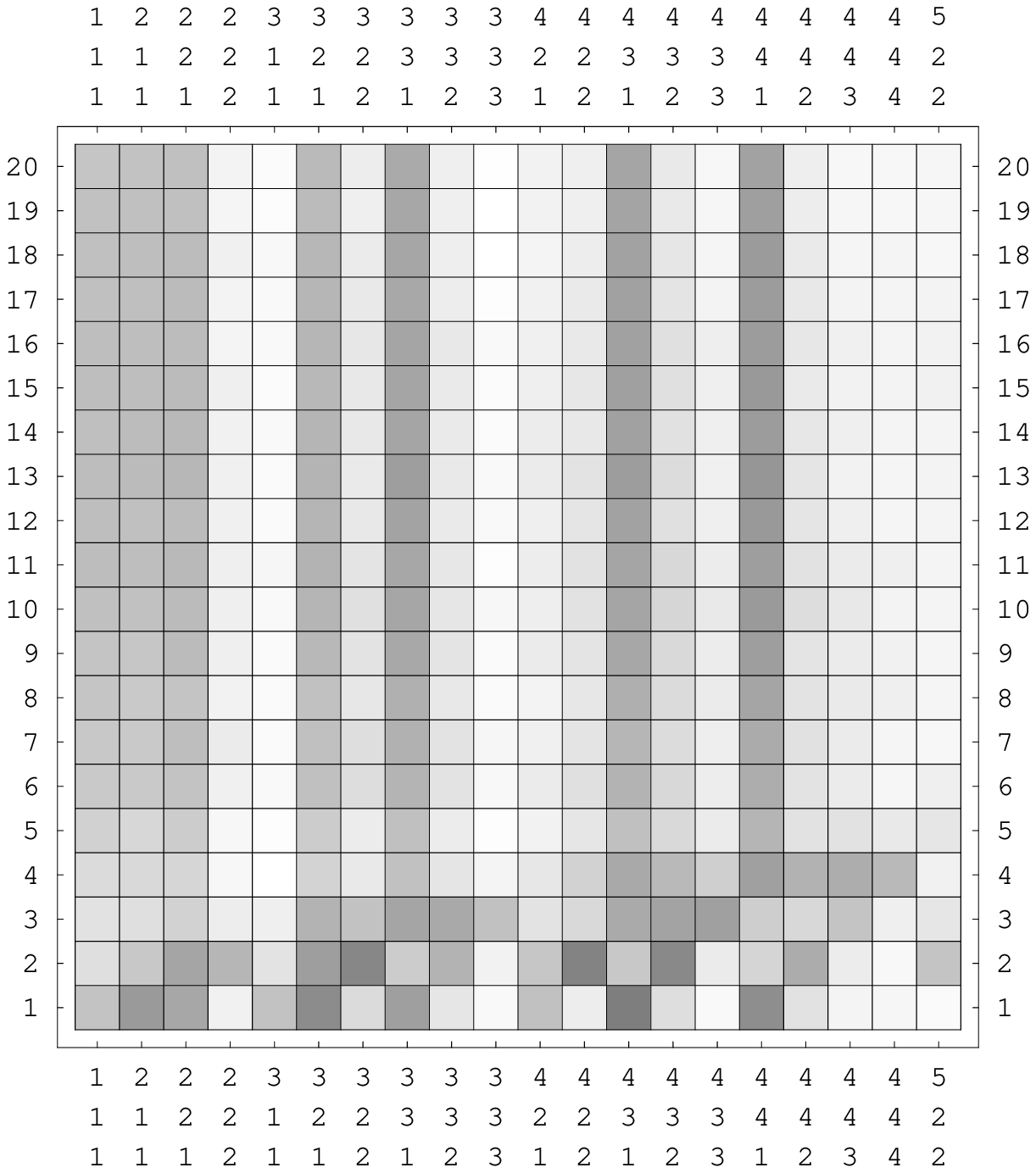}&
\includegraphics[width=0.49\textwidth]{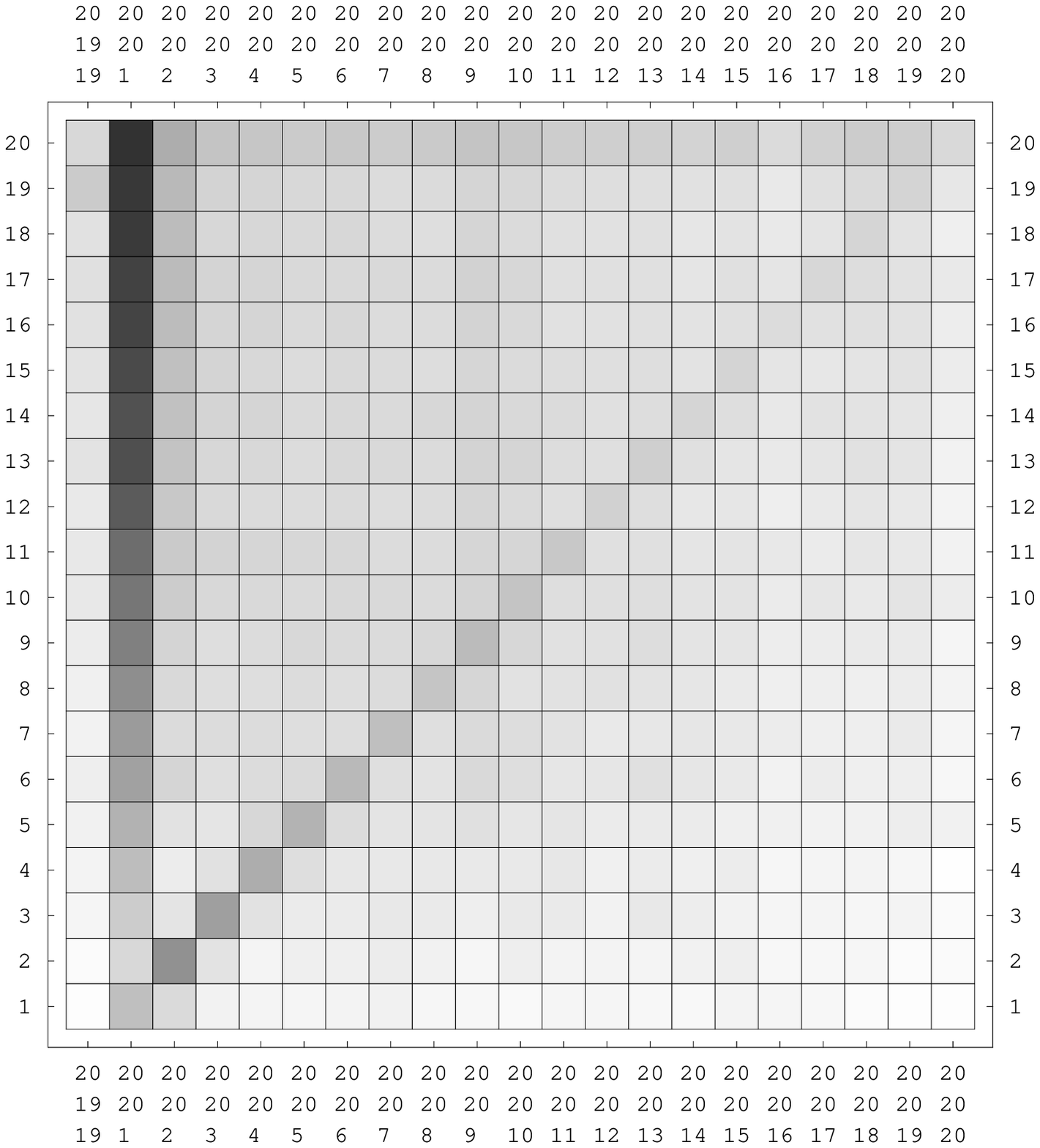}\\
\et
\caption{\label{densityPB} Mixed cross-correlation coefficients $r_{ij}^{PB}$ between main sample power spectrum and bispectrum. We show the $20$ largest scale triangles (left) and the $21$ smallest scale triangles (right) against all power spectrum bins. All numbers indicate wavenumbers in units of the $k$-bin width, $\Delta k\simeq0.015\kMpc$. The vertical bands are due to beat-coupling.}
%\bt{cc}
%\includegraphics[width=0.49\textwidth]{fig_ccPB_1_20.eps}&
%\includegraphics[width=0.49\textwidth]{fig_ccPB_995_1015.eps}\\
%\et
%\caption{\label{densityPB} Mixed cross-correlation coefficients $r_{ij}^{PB}$ between main sample power spectrum and bispectrum. We show the $20$ largest scale triangles (left) and the $21$ smallest scale triangles (right) against all power spectrum bins. All numbers indicate wavenumbers in units of the $k$-bin width, $\Delta k\simeq0.015\kMpc$. The vertical bands are due to beat-coupling.}
\end{center}
\end{figure*}

The difference in behavior is thus a reflection of non-linearity in the LRG bias, which creates additional non-Gaussianity. In fact, this is {\em expected} in standard scenarios of galaxies, since Eq.~(\ref{bhod}) naturally predicts that for galaxies that populate high-mass halos where $b_1 \simeq 2$, $b_2,b_3$ should be at least of order unity. However, we caution that, unlike the case of linear bias $b_1(m)$~\cite{Jing98,ShTo99,Sheth01,SeWa04,TWZZ05}, the expressions for non-linear bias parameters for halos $b_2(m),b_3(m)$ given by the peak-background split~\cite{MoWh96,ShTo99,Sheth01,SSHJ01} (also assumed by {\tt PTHalos}) have not been tested against current numerical simulations (see~\cite{JMW97} for early work). This is an important issue since the prediction is that $b_2,b_3$ are strong functions of halo mass for the range relevant for LRG galaxies (see e.g. Fig.~8 in~\cite{SSHJ01}), and small changes in the HOD parameters that leave the linear bias within observational bounds can change the non-linear bias parameters significantly. It is for this reason that we do not consider the bispectrum of LRG galaxies in this work, since its prediction has significant uncertainties. We are currently working on addressing these issues.

The nonlinearity in the bias relation for LRG galaxies implies that the power spectrum can only be reasonably well approximated by linear bias up to larger scales than in the main sample case. This is why we take $k_{\smax}=0.2\kMpc$ for LRG's, instead of $k_{\smax}=0.3\kMpc$ for the main sample. These are reasonable, though somewhat arbitrary values. In practice the allowed values of $k_{\smax}$ can be empirically tested by looking at higher-order correlations and looking for scale-dependence in the derived bias parameters~\cite{Scoccimarro:2000sp,Feldman:2000vk}.

\subsection{Bispectrum Covariance}
\label{secBScov}

In an analogous way to the power spectrum case, we can define the estimator for the bispectrum, \cite{Scoccimarro:1997st},
\bea\label{BSest}
\hat{B}(k_1,k_2,k_3) & \equiv & \frac{k_f^3}{V_B}\int_{k_1}\!\!\!\!d^3 q_1\int_{k_2}\!\!\!\!d^3 q_2\int_{k_3}\!\!\!\!d^3 q_3 \nonumber \\
& & \times\d_D(\qv_{123})\d_{\qv_1}\d_{\qv_2}\d_{\qv_3}, 
\eea
with
\bea
V_B & \equiv & \int_{k_1} \!\!\!\! d^3 q_1\int_{k_2} \!\!\!\! d^3 q_2 \int_{k_3} \!\!\!\! d^3 q_3 \,\delta_D(\qv_{123})\nonumber \\
& \simeq &  8\pi^2\ k_1 k_2 k_3\ \Delta k^3,
\eea
then the covariance between two triangle configurations (where $i$ and $j$ represents the triangles while $(i_1,i_2,i_3)$ and $(j_1,j_2,j_3)$ are the corresponding $k$-vectors triplets) is,
\bea\label{covBS}
C_{ij}^B & \equiv & \lan \delta B_i \delta B_j \ran \nonumber\\
& = & \d_{i_1j_1}\d_{i_2j_2}\d_{i_3j_3}\frac{k_f^3}{V_B(i)}P_{i_1}P_{i_2}P_{i_3} + {\rm cyc.}\nonumber\\
& & +\d_{i_1j_1}\frac{k_f^3}{V_B(i)V_B(j)}\intkqa\dots\!\!\intkqc\!\!\intkpb\!\!\intkpc\!\!\nonumber\\
& & \times\d_D(\qv_{123})\d_D(\qv_1\!\!+\!\!\pv_{23})B(\qv_1,\pv_2,\pv_3)B(\qv_1,\qv_2,\qv_3)\nonumber\\
& & + {\rm cyc.}\nonumber\\
& & +\d_{i_1j_1}\frac{k_f^3}{V_B(i)V_B(j)}\intkqa\dots\!\!\intkqc\!\!\intkpb\!\!\intkpc\!\!\nonumber\\
& & \times\d_D(\qv_{123})\d_D(\pv_{23}\!-\!\qv_1)P(q_1)T(\qv_2,\qv_3,\pv_2,\pv_3)\nonumber\\
& & +{\rm cyc.}\nonumber\\
& & +\frac{k_f^3}{V_B(i)V_B(j)}\intkqa\dots\!\!\intkqc\!\!\intkpa\dots\!\!\intkpc\nonumber\\
& & \times \d_D(\qv_{123})\d_D(\pv_{123}) T_{6}(\qv_1,\qv_2,\qv_3,\pv_1,\pv_2,\pv_3),
\eea
with $T_6(\kv_1,...,\kv_6)$ representing the 6-point connected correlation function in Fourier space. At large scales, the main contribution to the variance of the bispectrum is Gaussian and therefore
\beq
\Delta \hat{B}^2(k_1,k_2,k_3)\simeq s_B \frac{k_f^3}{V_B}\ P(k_1)P(k_2)P(k_3),
\label{Berror}
\eeq
with $s_B=6,2,1$ for equilateral, isosceles and general triangles, respectively. 

From the expression for $C_{ij}^B$ we see that the largest non-Gaussian contribution to the extra-diagonal elements of the bispectrum covariance matrix should arise in triangular configurations sharing two sides, with an extra factor when these are equal sides of isosceles triangles. Such large terms can be easily identified in Fig.~\ref{densityB} where we show the bispectrum cross-correlation coefficients. The value of the cross-correlation coefficients at the largest scales is given in Table~\ref{table_ccBS}. Note that, even at small scales, the bispectrum cross-correlation coefficients remain small, with values usually lower than $0.3$, typically quite smaller than in the power spectrum case.

\subsection{Mixed Covariance: Beat Coupling}
\label{secPBcov}

Given the estimators for the power spectrum and bispectrum defined in Eqs.~(\ref{PSest}) and~(\ref{BSest}), the mixed terms in the general covariance matrix are 
\bea
C_{ij}^{PB} & \equiv & \lan \delta P_i \delta B_j \ran =\nonumber\\
& \simeq & \d_{ij_1}\frac{2 k_f^3}{V_P(i)}P(k_i)B(k_{j_1},k_{j_2},k_{j_3})+ {\rm cyc.} \nonumber\\
& & +\frac{k_f^3}{V_P(i)V_B(j)}\intkqi\!\!\intkpa\dots\!\!\intkpc\!\!\nonumber\\
& & \times\d_D(\pv_{123}) T_{5}(\qv_1,-\qv_1,\pv_1,\pv_2,\pv_3),
\label{CPB}
\eea
where $T_{5}(\kv_1,\dots,\kv_5)$ stands for the 5-point connected correlation function in Fourier space. At large scales, the first term in Eq.~(\ref{CPB}) dominates, and moreover, this is expected to be an important contribution. To see this, compare its magnitude to the expected signal
\beq
\frac{\lan \delta P_i \delta B_j \ran}{P_i B_j}\simeq\frac{2 s_B k_f^3}{V_P(i)}\simeq\frac{s_B}{2\pi}\left(\frac{k_f}{k}\right)^2,
\eeq
which is comparable to the same ratio for the diagonal covariance of the power spectrum,
\beq
\frac{\Delta P^2_i}{P^2_i}\simeq\frac{2 k_f^3}{V_P(i)}\simeq \frac{1}{2\pi}\left(\frac{k_f}{k}\right)^2.
\eeq
Figure~\ref{densityPB} shows the cross-correlation coefficients between the first 20 (left) and last 21 (right) bispectrum configurations and all power spectrum bins. The terms just discussed correspond to the diagonal in the right panel, and a few of the elements in the bottom part of the left panel, where the power is calculated at one of the sides of the triangle. The value of the mixed cross-correlation coefficients at the largest scales is given in Table~\ref{table_ccPB}.

However, it is evident that there are significant correlations beyond these, for triangles which include the smallest value of $k$ as a side with {\em every} bin of the power spectrum. Indeed, Eq.~(\ref{CPB}) ignores important contributions that {\em dominate} the mixed covariance matrix. The reason is that so far we have ignored the effects of the window of the survey.

\begin{table}[b!]
\caption{\label{table_ccPB} Mixed cross-correlation coefficients between SDSS main sample bispectrum and power spectrum at the largest scales. Each triangular configuration is given in terms of the three vectors $k_1$, $k_2$, $k_3$ in units of the $k$-bin width, $\Delta k\simeq 0.015 \kMpc$.}
\begin{ruledtabular}
\begin{tabular}{l|llllllllll}
     & 1      & 2        & 3        & 4        & 5        & 6        & 7        & 8        & 9        & 10     \\  
\hline
1,1,1 & $0.24$ & $0.13$   & $0.12$   & $0.14$   & $0.18$   & $0.21$   & $0.22$   & $0.23$   & $0.23$   & $0.25$ \\
2,1,1 & $0.40$ & $0.22$   & $0.13$   & $0.15$   & $0.16$   & $0.21$   & $0.21$   & $0.21$   & $0.23$   & $0.24$ \\
2,2,1 & $0.34$ & $0.35$   & $0.18$   & $0.16$   & $0.20$   & $0.24$   & $0.26$   & $0.23$   & $0.27$   & $0.27$ \\
2,2,2 & $0.06$ & $0.29$   & $0.07$   & $0.03$   & $0.03$   & $0.06$   & $0.08$   & $0.05$   & $0.06$   & $0.06$ \\
3,1,1 & $0.24$ & $0.11$   & $0.06$   & $0.00$   & $0.01$   & $0.03$   & $0.02$   & $0.02$   & $0.02$   & $0.03$ \\
3,2,1 & $0.45$ & $0.38$   & $0.30$   & $0.17$   & $0.20$   & $0.25$   & $0.25$   & $0.25$   & $0.28$   & $0.29$ \\
3,2,2 & $0.14$ & $0.47$   & $0.24$   & $0.09$   & $0.08$   & $0.13$   & $0.13$   & $0.10$   & $0.11$   & $0.12$ \\
3,3,1 & $0.38$ & $0.20$   & $0.35$   & $0.24$   & $0.25$   & $0.29$   & $0.31$   & $0.31$   & $0.34$   & $0.35$ \\
3,3,2 & $0.10$ & $0.30$   & $0.34$   & $0.10$   & $0.07$   & $0.11$   & $0.11$   & $0.09$   & $0.11$   & $0.10$ \\
3,3,3 & $0.03$ & $0.05$   & $0.25$   & $0.04$   & $0.01$   & $0.03$   & $0.04$   & $0.02$   & $0.02$   & $0.03$ \\
\end{tabular}
\end{ruledtabular}
\end{table}

In a finite survey of size $\simeq L$, the uncertainty principle implies one cannot measure Fourier modes to a better accuracy than $\delta k \simeq \pi/L$, since two waves of frequency $k$ and $k\pm \delta k$ differ only by half an oscillation from one end to the other of the survey, {\it i.e.} there is not enough room inside the survey to tell them apart. This implies that in reality the power spectrum estimator in Eq.~(\ref{PSest}) written in terms of the {\em observed} Fourier modes will necessarily contain, due to the survey window, cross-terms in the underlying Fourier modes, written schematically as
\beq
\delta_\qv\ \delta_{-\qv+\epsilon},
\label{Pbeat}
\eeq
where $\epsilon \la \delta k$, apart from ``true power" contributions $\delta_\qv \delta_{-\qv}=|\delta_\qv|^2$. Although such terms do not contribute to the expectation value of the power, they do correlate very well with appropriate bispectrum configurations. Indeed, due to quadratic nonlinearities two nearby Fourier modes couple to the beat mode between them,
\beq
\delta_\qv\ \delta_{-\qv+\epsilon} \sim \delta_\qv\ F_2(-\qv,\epsilon)\, \delta_{-\qv}\, \delta_\epsilon,
\label{beatC}
\eeq
which means that these terms dominate the fluctuation in power at high wavenumbers where $q \gg \epsilon$, giving the non-intuitive result that the errors of the power spectrum in the nonlinear regime are dominated by the large-scale power~\cite{Hamilton:2005dx,Rimes:2005dz}. From Eq.~(\ref{beatC}) it follows that such terms cross-correlate very well with the bispectrum of isosceles triangles with one small side of the order of the survey window $\simeq \epsilon$,
\beq
\langle \delta_k\, \delta_{-k+\epsilon}\  \delta_p\, \delta_{-p}\, \delta_\epsilon \rangle \sim P(k) P(p)\, P(\epsilon).
\label{beatCeq25}\eeq
 Therefore, {\em for all} $k$ we expect power spectra to cross-correlate with bispectra of  ``narrow" isosceles triangles. These are the vertical features seen in Fig.~\ref{densityPB}.

Beat coupling implies that the whole power spectrum and the bispectrum of ``narrow" isosceles triangles fluctuate together depending on the large-scale power. As we shall see in section~\ref{likeBC} this has interesting implications for the likelihood analysis.

\section{The likelihood functions}
\label{secLikelihood}

We now consider a hypothetical joint analysis of large-scale structure (LSS) and cosmic microwave background (CMB) anisotropies. In order to be specific and illustrate the amount of information that we expect to extract in the very near future, we consider the first year WMAP data, the power spectrum and bispectrum of the SDSS main sample of galaxies, and also include the SDSS power spectrum of the luminous red galaxies (LRG).  The SDSS ``data" is obtained from the mock catalogs described in section~\ref{secBisp} and corresponds to the survey in its expected final form. In this section we describe the LSS and CMB likelihood functions that we use to derived the constraints discussed in the next sections.

\subsection{The LSS likelihood}

For simplicity we assume that the power spectrum and bispectrum estimators are Gaussian distributed. This is certainly a good approximation near the maximum wavenumber we include, but becomes worse at large scales, where only a few modes (for the power spectrum) or triangles (for the bispectrum) contribute. The deviations from Gaussian likelihood can be included by calculating the likelihood from the Monte Carlo pool of mock catalogs~\cite{Sco00b}. Ignoring the non-Gaussianity of the likelihood can lead to a biased estimation of parameters~\cite{Sco00b,SCJC00}. Since here we are only trying to understand the improvement brought by adding the bispectrum to the standard joint analysis of CMB and LSS, and most of the information added by the bispectrum is coming from scales small compared to the survey, our assumption should be safe. The combined power spectrum and bispectrum likelihood function is then
\beq
\ln{\mathcal L}=\ln{\mathcal L}_{P}+\ln{\mathcal L}_{B}+\ln{\mathcal L}_{PB},
\eeq
where
\beq\label{LP}
\ln{\mathcal L}_{P}=-\frac{1}{2}\sum_{i=1}^{N_k}\sum_{j=1}^{N_k}\delta P_i C^{-1}_{ij} \delta P_j,
\eeq 
\beq\label{LB}
\ln{\mathcal L}_{B}=-\frac{1}{2}\sum_{i=1}^{N_T}\sum_{j=1}^{N_T}\delta B_i C^{-1}_{ij} \delta B_j,
\eeq
and 
\beq\label{LPB}
\ln{\mathcal L}_{PB}=-\sum_{i=1}^{N_k}\sum_{j=1}^{N_T}\delta P_i C^{-1}_{ij} \delta B_j
\eeq
takes into account the mixed elements of the inverse covariance matrix $C^{-1}$. In Eqs.~(\ref{LP}-\ref{LPB}), the indices  $i$ and $j$ run over the bins in $k$-space for the power spectra, $N_k$ in all, as well as over the $N_T$ configurations included in the bispectrum analysis. Also, $\delta P\equiv P_s-P_s^*$ and $\delta B \equiv B_s - B_s^*$, where $P_s\equiv P_s(\pv;k)$ and $B_s\equiv B_s(\pv;k_1,k_2,k_3)$ are the redshift space galaxy power spectrum and bispectrum as a function of the parameters $\pv$ while $P_s^*\equiv P_s(\pv^*;k)$ and $B_s^*\equiv B_s(\pv^*;k_1,k_2,k_3)$ are the redshift space galaxy power spectrum and bispectrum of the fiducial model (with parameters $\pv^*$). 

 In the most general case we consider
\beq\label{pvector} 
\pv=(\tau,A_s,\omega_d,\omega_b,\Omega_{\Lambda},n_s,w),
\eeq
defined as the reionization optical depth, $\tau$, the primordial amplitude of scalar fluctuations, $A_s$, the physical dark matter density, $\omega_d\equiv\Omega_d h^2$, the physical baryon density, $\omega_b\equiv\Omega_b h^2$, the dark energy density, $\Omega_{\Lambda}$, the scalar spectral index, $n_s$ and the dark energy equation of state parameter, $w=p_{\Lambda}/\rho_{\Lambda}$. The bias parameters include the main sample linear and quadratic bias, $b_1$ and $b_2$, and the LRG linear bias $b_1^{LRG}$. 

The covariance matrices are calculated at maximum likelihood from our mock catalogs, that is, we do not include a possible dependence on parameters to be estimated.  A simple, approximate, check of such dependence on the bias parameters did not yield appreciable differences in the final results we present in section~\ref{secResults}. 

When we study below results from the power spectrum or bispectrum individually the inverse matrix $C^{-1}$ in Eq.~(\ref{LP}) or~(\ref{LB}) will be replaced by the inverse of the individual matrix $C_{ij}^P\equiv\lan\d P_i\d P_j\ran$ or $C_{ij}^B\equiv\lan\d B_i\d B_j\ran$. We will study as well the case of combining the two statistics {\em without} taking into account their mixed covariance, in which case also only $C^P$ and $C^B$ will be needed.

The likelihood function for the LRG power spectrum is equivalent to Eq.~(\ref{LP}) and the corresponding covariance matrix is independently determined from the LRG mock catalogues. Since the mean redshift of the LRG sample is $z\simeq0.35$ compared to $z\simeq0.1$ for the main sample, with little overlap, we assume the two samples are independent.

\subsubsection{Power Spectrum}
\label{secLHps}

Deviations from the fiducial {\it redshift space} power spectrum monopole $P^*_s$, as a function of the parameters $\pv$, are modeled in the following way
\beq\label{MSpowerspectrum}
P_s(\pv;k)= b_1^2~f_s^P(\pv)f(\pv;k)~P^*_s(k),
\eeq
where $P^*_s$ is the power spectrum measured from our mocks catalogs in redshift space and where
\bea
f(\pv;k) & \equiv & \frac{A}{A^*}\left[\frac{T(\pv;k)}{T^*(k)}\right]^2\left[\frac{D(\pv)/\Omega_m}{D^*/\Omega_m^*}\right]^2\nonumber\\
& & \times\left(\frac{k}{k_P}\right)^{n_s-n_s^*}\frac{h^{n_s}}{(h^*)^{n_s^*}},
\eea
$T(\pv;k)$ being the transfer function, $D_+(\pv)$ the growth factor and $k_P=0.05~Mpc^{-1}$ the pivot point corresponding to the scale whose power is unaffected by varying the spectral index, and
\beq%\label{def_fsP}
f_s^P(\pv)\equiv\frac{a_0(\beta)}{a_0(\beta^*)}
\eeq
is the redshift-space correction where 
\beq
a_0=1+\frac{2}{3}\beta+\frac{1}{5}\beta^2,
\label{Kfactor}
\eeq
with $\beta\simeq\Omega_m^{5/9}/b_1$, corresponds to the power spectrum monopole~\cite{Kaiser87}. Note that we use Eq.~(\ref{Kfactor}) only to model {\em deviations} from our fiducial cosmology assumed in the mock catalogs that include nonlinear effects from the redshift-space mapping. We assume a fiducial model that is unbiased, $b_1=1$ and $b_2=0$. The likelihood function for the LRG power spectrum is computed in the same way, except for the fiducial value of the LRG linear bias parameter $b_1^{LRG}=2.17$.

Note that, since redshift distorsions break the statistical isotropy expected in real-space, the redshift-space power spectrum is a function of the direction as well as the magnitude of the wavevector $\kv$. In this work we include only the monopole of the redshift-space power spectrum, {\it i.e.} the average over the angle formed by $\kv$ and the line of sight. In principle one can take advantage of the angle-dependence by measuring the quadrupole term in the Legendre polynomial expansion and obtain a better determination of the $\beta$ parameter, further strengthening the constraints presented in section~\ref{secResults} below. 

We calculate the transfer functions from CMBFAST~\cite{Seljak:1996is}, computing the value of $T(\pv;k)$ for every value on a limited grid in parameter space, then interpolating over the final parameter grid for each value of the wavenumber $k$ involved in the analysis.

For the growth factor $D(\Omega_m,w)$ we take advantage of the fitting formula provided in~\cite{Linder:2005in}. This is given by
\beq
D(a)=a\exp\left\{\int_0^a \de \ln a \left[\Omega_m(a)^\gamma-1\right]\right\},
\eeq
with $\gamma=0.55+0.05(1+w)$ and where $a$ is the cosmological scale factor.

\subsubsection{Bispectrum}

We describe the deviations of the bispectrum from the fiducial model of the mock catalogs by means of Eulerian perturbation theory. Since we are averaging the redshift-space bispectrum over triangles with all possible orientations, similarly to the power spectrum case, we only need the {\em monopole} term in a Legendre expansion. We will consider the following approximation, see Eqs.~(20-28) in~\cite{Scoccimarro:1999ed}
\bea
B_s^{(0)} & \simeq & a_0^B(\beta)\frac{2}{b_1} F_2(\kv_1,\kv_2) P_g(k_1)P_g(k_2) \nonumber \\
& & +a_0^B(\beta) \frac{b_2}{b_1^2} P_g(k_1)P_g(k_2)+ {\rm cyc.},
\eea
where $P_g(k)=b_1^2 P(k)$ is the galaxy power spectrum, and
\beq
a_0^B(\beta)\equiv 1+\frac{2}{3}\beta+\frac{1}{9}\beta^2,
\eeq
describes the bispectrum monopole redshift space correction, obtained from Eq.(24) and (28)  in~\cite{Scoccimarro:1999ed} by averaging over the angle between $\kv_1$ and $\kv_2$ and dropping the dependence on the second-order velocity kernel and velocity dispersion which should partially cancel at large scales, to approximate the configuration dependence found in simulations for the redshift-space bispectrum in~\cite{Scoccimarro:1999ed}. 

To compute the dependence of the bispectrum on cosmological parameters we will therefore use
\bea
B_s(\pv) & = & f_s^B(\pv,)~b_1^3~{\mathcal B}~\frac{B_s^*}{{\mathcal B}^*}\nonumber\\
& & + f_s^B(\pv)~b_2~b_1^2~{\mathcal S},
\eea
where $B_s^*$ is the redshift-space bispectrum measured from the mock catalogs and
\bea
{\mathcal B} & = & 2 F_2(\kv_1,\kv_2) P_s^*(k_1) f_1 P_s^*(k_2) f_2+{\rm cyc.},\\
\bar{\mathcal S} & = & P_s^*(k_1) f_1 P_s^*(k_2) f_2 +{\rm cyc.},
\eea
while
\beq
f_s^B(\pv,\pv^*)\equiv\frac{a_0^B(\beta)}{a_0^B(\beta^*)},
\eeq
with $\beta^*$ as fiducial $\beta$ and $f_1\equiv f(\pv;k_1)$ as defined above. 

\subsubsection{Inverting the Covariance Matrix}

The values of the entries of the complete covariance matrix $C_{ij}\equiv\lan X_iX_j\ran$ with $X_i=P_i,B_i$ span several orders of magnitude and thus a direct computation of its inverse is susceptible to numerical instabilities. We therefore ``normalize'' the covariance matrix by factoring out in the $X={P,B}$ vector the power spectrum and bispectrum predicted by linear theory and Eulerian PT, respectively. The resultant entries for this ``normalized'' covariance matrix are therefore all of order unity. Still, by performing a singular value decomposition (SVD) one can notice a poor determination of a few singular values, about 17 out of 1035 in the complete, $k_{\smax}^{MS}=0.3\kMpc$ case, indicating that 6000 mock catalogs is enough to determine most of the elements except for a small fraction. In the final analysis we compute the inverse by means of its SVD inverse by dropping these few singular values, assuming this might be a sign of a not optimal determination of the matrix $C_{ij}$. By doing this we make a conservative choice since the operation amounts to discard some of the potential information contained in the covariance matrix. Therefore our final error bars increase slightly. In computing the inverse of the individual power spectrum and bispectrum covariance matrices no such limitation is needed.

\subsection{The CMB likelihood}
\label{subsectionCMB}

To combine the results of the LSS likelihood analysis with CMB data as measured by WMAP 1-year, we need to compute the CMB anisotropies power spectrum and its corresponding likelihood for each model in our grid. This procedure is computationally expensive: version 4.5.1 of the CMBFAST code~\cite{Seljak:1996is} takes about $30$ seconds per model, and the WMAP likelihood code for the first year data release~\cite{Verde:2003ey} takes $2-4$ seconds.

A possible approach to reduce the computing time is investigated in~\cite{Sandvik:2003ii}, where a polynomial approximation to the multidimensional log-likelihood is computed, allowing for an evaluation of the likelihood of each model in tenths of a second. However, their available code, CMBFIT 1.0, does not include the dark energy equation of state parameter, $w$. Although calculating such a polynomial fit still requires sampling the likelihood surface, it has to be done only once, thus reducing enormously the computational time needed for any ulterior likelihood analysis. 

Motivated by that idea, we compute a polynomial fit to the CMB likelihood function based on the $7$ parameters in Eq.~(\ref{pvector}). We use uniform priors in the following ranges: $0 \leq \tau \leq 0.3$, $0.018 \leq \omega_b \leq 0.028$,$-2 \leq w \leq 0.2$,$0.5 \leq A_s \leq 1.4$,$0.4 \leq \Omega_{\Lambda} \leq 0.9$,$0.08 \leq \omega_d \leq 0.22$,$0.8 \leq n_s \leq 1.1$.

We compute the likelihood on a homogeneous grid with 15 points per dimension for the parameters $(\omega_d,\omega_b,\Omega_{\Lambda},n_s,w)$, and 30 points for $(\tau,A_s)$. To speed up the calculation, we divided the problem in two steps: firstly, we computed the likelihood on the grid approximating the dependence  of the power spectrum on $\tau$ with the multiplicative factor $e^{-2\tau}$. Out of these $7\times 10^8$ approximate likelihood values we selected a connected subset of $263,022$ models containing the maximum and defined by a threshold chosen to be 10 orders of magnitude smaller than the maximum. We then recomputed the likelihood for the reduced subset with the correct $\tau$ dependence.

We fitted a 4th order polynomial to the log-likelihood surface spanned by our reduced dataset using a weighted least squares method. We weighted the fitting error of each model with its likelihood to counterbalance the fact that our grid was relatively coarse and there were many more low likelihood models than high likelihood ones.
The covariance matrix of our 7-dimensional reduced set of models is given by $C_{ij}\equiv\langle p_i p_j \rangle-\langle p_i \rangle \langle p_j \rangle$, with $\langle p_i \rangle \equiv \sum p_i \, {\mathcal L}({\bf p})$. In order to improve the numerical behavior of the fitting algorithm we first changed from ${\bf p}-$space to the variables ${\bf z}$ with zero mean and unit covariance defined as
\beq
\zv \equiv {\bf E}(\pv-\langle\pv\rangle),
\eeq
where the rows of ${\bf E}$ were defined as the eigenvectors of the covariance matrix ${\bf C}$ divided by the square root of their corresponding eigenvalues, {\it i.e.} such that ${\bf E C} {\bf E}^t = {\bf I}$, and thus $\langle {\bf z} {\bf  z}^t \rangle = {\bf I}$, \cite{Sandvik:2003ii}. The 4th order polynomial, containing $M=330$ terms, can be written in terms of the new variables as
\begin{eqnarray}
y\equiv \ln {\mathcal L}=q_0+\sum_{i_1=1}^7 z_{i_1}\left\{\,q_1^{i_1}+\sum^7_{i_2=i_1} z_{i_2}\left[ \,q_2^{i_1 i_2}+\right.\right. \nonumber \\  
\left. \left. \sum_{i_3=i_2}^7 z_{i_3}\left( q_3^{i_1 i_2 i_3}+\sum^{7}_{i_4=i_3}q_4^{i_1 i_2 i_3 i_4} z_{i_4} \right) \right] \right\}. \ \ \ \
\label{yeq}
\end{eqnarray}
\begin{figure}[t]
\begin{center}
{\includegraphics[width=0.48\textwidth]{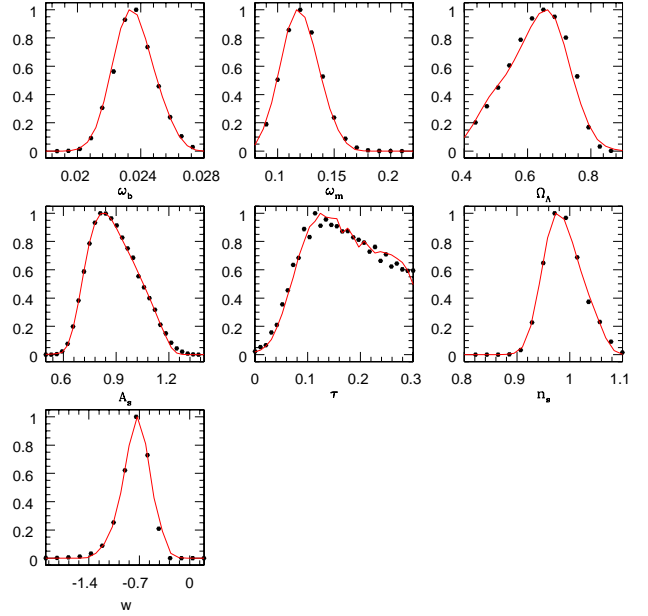}}
\caption{WMAP TT+TE marginalized likelihoods for the $7$ cosmological parameters in our wCDM model. Dots correspond to the marginalization over the reduced dataset in the grid (263,022 points) while the solid line was obtained using the {\it weighted} polynomial fit.}
\label{figfitdata}
\end{center}
\end{figure}
In order to make this expression compact we arranged all possible products of $z_i$'s (up to the 4th power) into an $M$-dimensional array ${\bf x}$,
\begin{equation}
\xv=\{1,z_1,\dots,z_7,z_1 z_1,z_1 z_2,\dots,z_7^4\},
\end{equation}
and the corresponding coefficients $q$ into, 
\begin{equation}
\qv=\{q_0,q_1^1,\dots,q_1^7,q_2^{11},\dots,q_4^{7 7 7 7}\},
\end{equation}
so that Eq.~(\ref{yeq}) could be casted as 
\begin{equation}
y=\xv \cdot \qv.
\end{equation}
Next, we arranged our $N=263,022$ datapoints, written in the $\xv$ format, into an ($N\times M$) matrix $\Xv$ and their corresponding likelihoods into an $N$-dimensional vector $\yv$. Therefore, the weighted error of the polynomial fit was given by,
\begin{equation}
\epsilon\equiv\left[\frac{1}{M}\sum_{i=1}^M w_i (y_i-\sum_{j=1}^M X_{ij} q_j)^2\right]^{1/2}
\end{equation}
where $w_i={\mathcal L}(\pv_i)$. The coefficients $q_i$ were then chosen such as to minimize $\epsilon$,
\begin{equation}
\qv=(\tilde{\Xv}^t \Xv )^{-1} \tilde{\Xv}^t \yv,
\end{equation}
where we defined $\tilde{\rm X}_{ij}=w_i {\rm X}_{ij}$. 

In order to avoid unphysical likelihood values due to polynomial artifacts in low confidence regions that were poorly sampled, we replaced the polynomial fit by a simple Gaussian for $z=|\zv|$ outside the $2$-sigma level, given  that the likelihood distribution has a spherically symmetric tail for $z>2.5$, well approximated by ${\mathcal L} \propto {\rm e}^{ -z^2/2}$.

To test our weighted fit we compared it, for ${\rm w}=-1$, against CMBFIT for a 6-parameter $\Lambda$CDM model ($\tau,A_s,\omega_d,\omega_b,\Omega_{\Lambda},n_s$), finding very good agreement. Another estimator of the goodness of the fit was the rms fitting error $\langle\Delta \ln\mathcal{L}\rangle$. We found that $\langle\Delta \ln\mathcal{L}\rangle=0.29$ for our fitting dataset of $263,022$ models. Furthermore, we made a Monte Carlo Markov Chain test of 3000 models, which yielded $\langle\Delta \ln\mathcal{L}\rangle=0.32$. These errors are similar to those reported in~\cite{Sandvik:2003ii} for the 7-parameter models of CMBFIT.
\begin{table}[t]
\caption{\label{fiducialvalues6P} Fiducial values for the cosmological and bias parameters assumed.}
\begin{ruledtabular}
\begin{tabular}{llcc}
\multicolumn{2}{l}{Parameter}                         &  Fiducial value \\\hline
$\omega_d$        & physical dark matter density      & $0.1222$        \\
$\omega_b$        & physical baryon density           & $0.0232$        \\
$\Omega_\Lambda$  & dark energy density               & $0.699$         \\
$n_s$             & scalar spectral index             & $0.977$         \\
$A_s$               & scalar fluctuation amplitude      & $0.81$          \\
$w$               & dark energy equation of state     & $-1$            \\
$\tau$            & reionization optical depth        & $0.124$         \\
$b_1$             & main sample linear galaxy bias    & $1$             \\
$b_2$             & main sample quadratic galaxy bias & $0$             \\
$b_1^{LRG}$       & LRG linear galaxy bias            & $2.17$          \\\hline
\multicolumn{2}{l}{Derived parameters}                &  Fiducial value \\\hline
$\sigma_8$        & galaxy fluctuation amplitude      & $0.917$         \\
$\Omega_m$        & matter density                    & $0.301$         \\
$\Omega_b$        & baryon density                    & $0.048$         \\
$h$               & Hubble parameter                  & $0.695$         \\
\end{tabular}
\end{ruledtabular}
\end{table}

Figure~\ref{figfitdata} shows the first year WMAP TT+TE marginalized likelihoods for the case of the $7$-parameter $w$CDM models, Eq.~(\ref{pvector}), obtained by a grid marginalization over the reduced dataset (dots) versus the ones using the weighted polynomial fit (solid line). This shows the procedure described above is robust enough for our purposes.

\section{Results}
\label{secResults}
\begin{table*}[t!]
\caption{\label{tab_results_9P} $\Lambda$CDM models: expected marginalized errors ($68\%$ CL) for WMAP1 (temperature and polarization, W) combined with the SDSS main sample power spectrum (P) and bispectrum (B) and with the LRG power spectrum (P$_L$). The percentage in parenthesis indicates the improvement over the analysis including the main sample power spectrum alone (W+P), numbers in bold indicate errors down by at least $1.5$. In brackets we quote the W+P+B errors obtained by ignoring the mixed power spectrum - bispectrum covariance.}
\begin{ruledtabular}
\begin{tabular}{l|lllll|ll}
  & W+P & W+B & W+P+B & W+P+B (no mix. cov.) & W+P+B+P$_L$ & W+P+P$_L$ & W+P$_L$ \\
\hline\hline
 & \multicolumn{5}{l}{$k_{\smax}^{MS}=0.2 \kMpc$} & \multicolumn{2}{l}{$k_{\smax}^{LRG}=0.2 \kMpc$}\\\hline\hline
$\Delta\omega_d$      &$0.0035$ &$0.0041$~($-15\%$)&$0.0031$~($13\%$) &[~$0.0030$~($17\%$)~] &$0.0025$~($40\%$) &$0.0026$~($35\%$) &$0.0029$~($21\%$)  \\\hline
$\Delta\omega_b$      &$0.00093$&$0.00098$~($-5\%$)&$0.00078$~($19\%$)&[~$0.00082$~($13\%$)~]&$0.00074$~($26\%$)&$0.00081$~($15\%$)&$0.00087$~($7\%$)  \\\hline
$\Delta\Omega_\Lambda$&$0.0133$ &$0.0113$~($18\%$) &${\bf 0.0085}$~($56\%$) &[~$0.0085$~($56\%$)~] &${\bf 0.0063}$~($111\%$) &${\bf 0.0078}$~($70\%$) &$0.0091$~($46\%$)  \\\hline
$\Delta n_s$          &$0.022$  &$0.024$~($-8\%$)  &$0.0158$~($39\%$) &[~$0.0176$~($25\%$)~] &${\bf 0.0140}$~($57\%$) &$0.0171$~($28\%$) &$0.020$~($10\%$)    \\\hline
$\Delta A_s$            &$0.091$  &$0.094$~($-3\%$)  &$0.064$~($42\%$)  &[~$0.074$~($23\%$)~]  &$0.062$~($47\%$)  &$0.078$~($17\%$)  &$0.085$~($7\%$)    \\\hline
$\Delta \tau$         &$0.052$  &$0.053$~($-2\%$)  &$0.039$~($33\%$)  &[~$0.044$~($18\%$)~]  &$0.038$~($37\%$)  &$0.047$~($11\%$)  &$0.049$~($6\%$)    \\\hline
$\Delta b_1$          &$0.086$  &$0.113$~($-24\%$) &$0.060$~($43\%$)  &[~$0.074$~($16\%$)~]  &${\bf 0.054}$~($59\%$)  &$0.070$~($23\%$)  & -                 \\\hline
$\Delta b_2$          & -       &$0.069$           &$0.054$           &[~$0.062$~]           &$0.051$           & -                & -                 \\\hline
$\Delta b_1^{LRG}$    & -       & -                & -                & -                    &$0.099$           &$0.125$           &$0.137$            \\\hline
\hline
$\Delta \sigma_8$     &$0.068$  &$0.074$~($-8\%$)  &$0.047$~($45\%$)  &[~$0.054$~($26\%$)~]  &${\bf 0.043}$~($58\%$)  &$0.054$~($26\%$)  &$0.061$~($11\%$)   \\\hline
$\Delta h$            &$0.0152$ &$0.0140$~($9\%$)  &${\bf 0.0101}$~($50\%$) &[~$0.0108$~($41\%$)~] &${\bf 0.0087}$~($75\%$) &$0.0106$~($43\%$) &$0.0127$~($19\%$)  \\\hline
$\Delta \Omega_b$     &$0.00151$&$0.00141$~($7\%$) &$0.00124$~($22\%$)&[~$0.00124$~($22\%$)~]&$0.00111$~($36\%$)&$0.00117$~($29\%$)&$0.00127$~($19\%$) \\\hline
\hline
 & \multicolumn{5}{l}{$k_{\smax}^{MS}=0.3 \kMpc$} & \multicolumn{2}{l}{$k_{\smax}^{LRG}=0.2 \kMpc$}\\\hline\hline
$\Delta\omega_d$      &$0.0033$ &$0.0031$~($6\%$)  &$0.0029$~($14\%$) &[~$0.0026$~($27\%$)~] &$0.0024$~($37\%$)   & $0.0026$~($27\%$)  & \\\hline
$\Delta\omega_b$      &$0.00090$&$0.00083$~($8\%$) &$0.00073$~($23\%$)&[~$0.00072$~($25\%$)~]&$0.00070$~($28\%$)  & $0.00080$~($12\%$) & \\\hline
$\Delta\Omega_\Lambda$&$0.0112$ &${\bf 0.0065}$~($72\%$) &${\bf 0.0063}$~($78\%$) &[~$0.0057$~($96\%$)~] &${\bf 0.0052}$~($115\%$)   & ${\bf 0.0073}$~($53\%$)  & \\\hline
$\Delta n_s$          &$0.021$  &$0.018$~($17\%$)  &${\bf 0.014}$~($50\%$)  &[~$0.013$~($61\%$)~]  &${\bf 0.012}$~($75\%$)    & $0.016$~($31\%$)   & \\\hline
$\Delta A_s$            &$0.087$  &$0.081$~($7\%$)   &${\bf 0.053}$~($64\%$)  &[~$0.063$~($38\%$)~]  &${\bf 0.052}$~($67\%$)    & $0.077$~($13\%$)   & \\\hline
$\Delta \tau$         &$0.050$  &$0.047$~($6\%$)   &${\bf 0.033}$~($52\%$)  &[~$0.039$~($28\%$)~]  &${\bf 0.033}$~($52\%$)    & $0.046$~($9\%$)    & \\\hline
$\Delta b_1$          &$0.081$  &$0.094$~($-14\%$) &${\bf 0.051}$~($59\%$)  &[~$0.060$~($35\%$)~]  &${\bf 0.046}$~($76\%$)    & $0.068$~($19\%$)   & \\\hline
$\Delta b_2$          & -       &$0.045$           &$0.041$           &[~$0.044$~]           &$0.040$             & -                  & \\\hline
$\Delta b_1^{LRG}$    & -       & -                & -                & -                    &$0.084$             & $0.123$            & \\\hline\hline
$\Delta \sigma_8$     &$0.064$  &$0.059$~($8\%$)   &${\bf 0.037}$~($73\%$)  &[~$0.041$~($56\%$)~]  &${\bf 0.034}$~($88\%$)    & $0.053$~($21\%$)   & \\\hline
$\Delta h$            &$0.0132$ &$0.0095$~($39\%$) &${\bf 0.0082}$~($61\%$) &[~$0.0080$~($65\%$)~] &${\bf 0.0082}$~($61\%$)   & $0.0101$~($31\%$)  & \\\hline
$\Delta \Omega_b$     &$0.00138$&$0.00112$~($23\%$)&$0.00111$~($24\%$)&[~$0.00106$~($30\%$)~]&$0.00104$~($33\%$)  & $0.00115$~($20\%$) & \\
\end{tabular}
\end{ruledtabular}
\end{table*}

In this section we present the results of the likelihood analysis in two classes of flat cosmological models. The first, section~\ref{secLCDM}, corresponds to $\Lambda$CDM models depending on six cosmological plus three bias parameters: the density parameters $\omega_d$, $\omega_b$, $\Omega_\Lambda$, the spectral index $n_s$, the fluctuations amplitude $A_s$, the reionization optical depth $\tau$ plus the linear and quadratic galaxy bias coefficients $b_1$ and $b_2$ for the main sample and the linear bias for the LRG sample, $b_1^{LRG}$. In the second class, denoted as $w$CDM models, section~\ref{secwCDM}, we allow for a dark energy equation of state parametrized by the ratio of pressure to energy density $w$, assumed to be constant.

We include the temperature and polarization WMAP 1-year likelihood by means of the interpolation fit described in section~\ref{subsectionCMB} (for an update to the 3-year data, see the Appendix). We introduce here a flat prior on $\tau$ by limiting its values from zero to $0.3$. The difference with the case of $\tau$ taking values up to $0.8$ is negligible (tested for $w=-1$) and, most importantly, such a prior is more than justified by the three-year WMAP data which favors values of $\tau$ close to $0.1$~\cite{Spergel:2003cb,Page06}.

The fiducial values chosen for the present analysis are given in Table~\ref{fiducialvalues6P}. Note that they do not coincide with the maximum likelihood values obtained from the WMAP data alone, rather they correspond to those obtained for the WMAP+SDSS power spectrum 6 parameters case in~\cite{Tegmark:2003ud}. These values are only relevant in the sense that they determine the point in parameter space about which we compute the errors. As long as this point is realistic, the results we present should be insensitive to their precise values.

\subsection{$\Lambda$CDM models}
\label{secLCDM}
\begin{figure*}[t]
\begin{center}
\includegraphics[width=0.98\textwidth~]{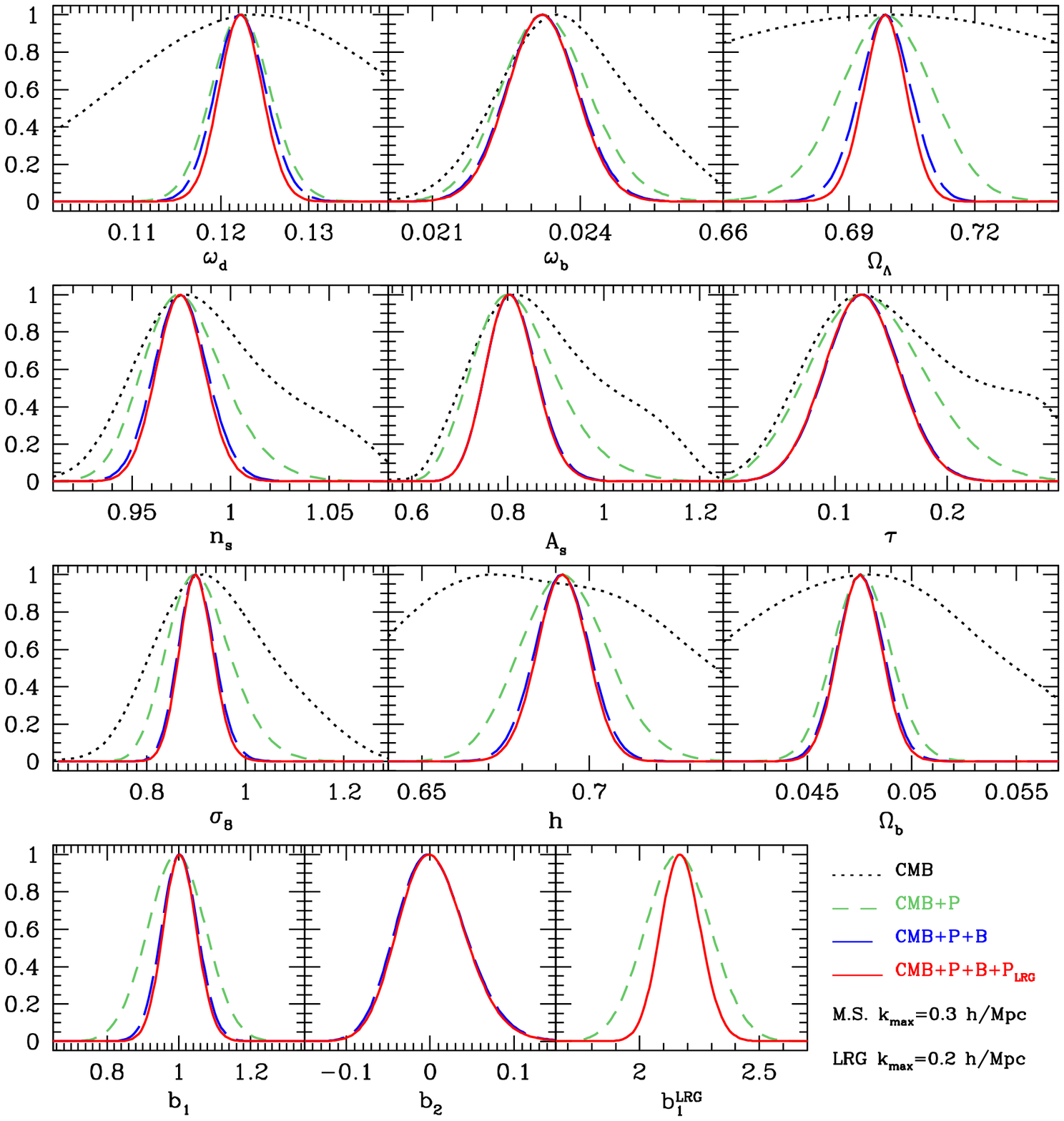}
\caption{\label{fig9Pk3} Marginalized likelihood functions for the $\Lambda$CDM models assuming $k_{\smax}^{MS}=0.3 \kMpc$. For the LRG linear bias parameter $b_1^{LRG}$ only, the dashed line denotes the likelihood obtained from WMAP1 plus {\it LRG} power spectrum instead of main sample power spectrum.}
\end{center}
\end{figure*}

We present now the expected errors on cosmological parameters from an analysis that considers different combinations of the main sample power spectrum and bispectrum and the power spectrum of the LRG sample with WMAP CMB data. We restrict here to the case of $\Lambda$CDM models, {\it i.e.} $w=-1$. The results for the 1-$\sigma$ marginalized uncertainties are given in Table~\ref{tab_results_9P} where we show, for clarity, the average between upper and lower limits. 

To see more clearly the benefits brought by using different statistics, in parenthesis we indicate the fractional improvement over the WMAP plus main sample power spectrum case (W+P), defined as 
\beq
{\rm improvement~factor} = \frac{\Delta_{\rm W+P}}{\Delta}-1,
\label{improve}
\eeq
so a $50\%$ ($100\%$) improvement corresponds to reducing the errors by a factor of $1.5$ (2).

\begin{figure*}[t]
\begin{center}
\includegraphics[width=0.98\textwidth~]{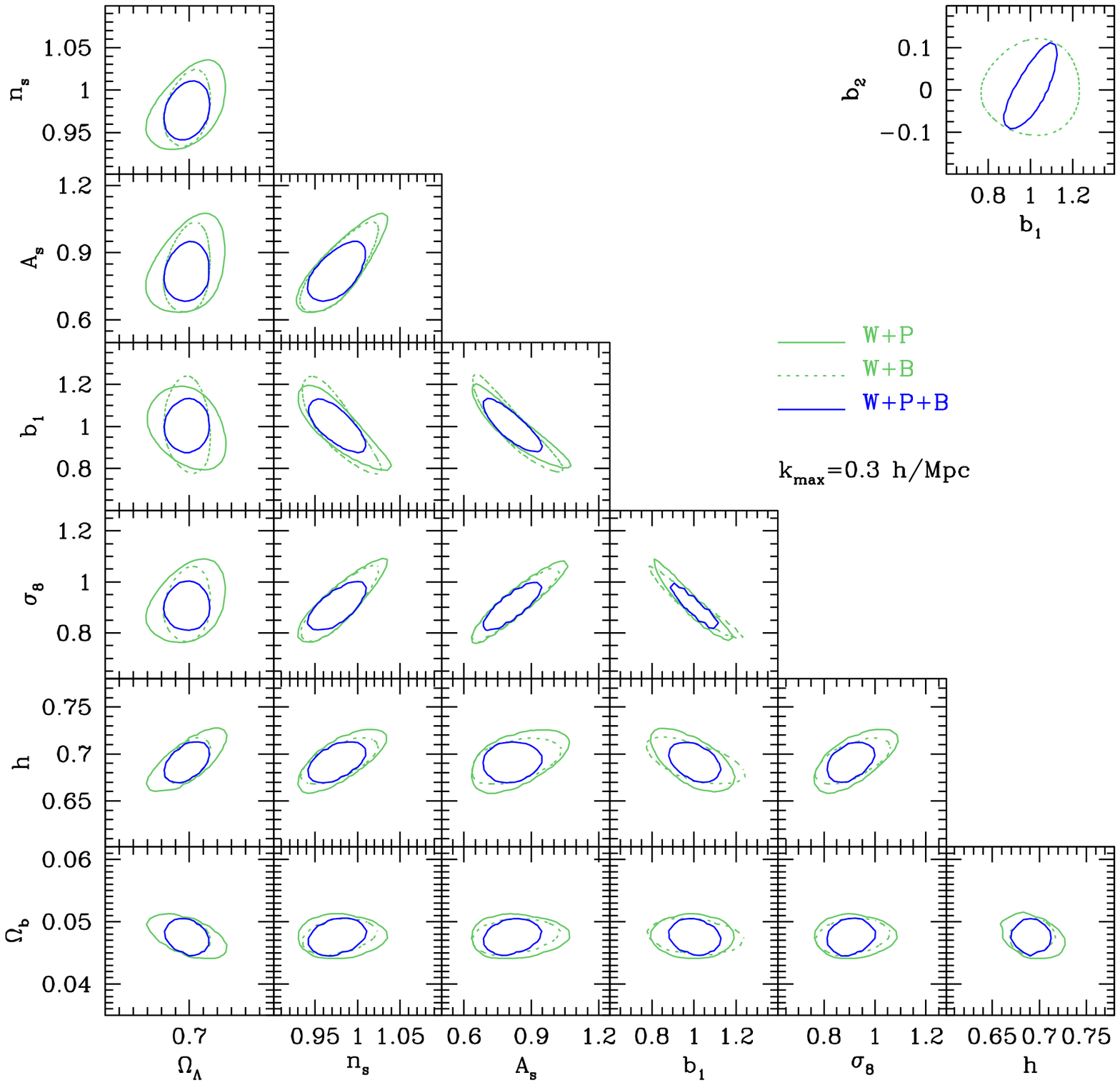}
\caption{\label{fig_9P_k3_contours} Marginalized $95\%$ contour plots for pairs of cosmological parameters in the $\Lambda$CDM models assuming $k_{\smax}^{MS}=0.3 \kMpc$.}
\end{center}
\end{figure*}

The first two columns in Table~\ref{tab_results_9P} show that analyzing the power spectrum and bispectrum separately can provide similar constraints (with the bispectrum determining an extra parameter, $b_2$). This can provide important consistency checks, as the Gaussian and non-Gaussian properties of galaxy clustering must yield consistent results.

\begingroup
\begin{table*}[t!]
\caption{\label{results9PavgPvsB} Comparison between W+P and W+P+B with and without galaxy bias assumptions. Includes W+P with a prior on $b_1$ to reproduce the error from W+P+B (left) and a case where the bias parameters are fixed (right). The percentage in parenthesis indicate the improvement over the respective W+P result. Assumes a $\Lambda$CDM cosmology and $k_{\smax}^{MS}=0.3\kMpc$.}
\begin{ruledtabular}
\begin{tabular}{l|llll|ll}
                      & W+P     & W+P+B            &W+P+ $b_1$ prior   &W+P+$b_1$ prior ($Q$)&W+P fixed bias&W+P+B fixed bias \\\hline\hline
$\Delta\omega_d$      &$0.0033$ &$0.0029$~($14\%$) &$0.0032$~($3\%$)   &$0.0031$~($6\%$)     &$0.0031$      &$0.0026$~($19\%$) \\\hline
$\Delta\omega_b$      &$0.00090$&$0.00073$~($23\%$)&$0.00078$~($15\%$) &$0.00073$~($23\%$)    &$0.00068$     &$0.00065$~($5\%$) \\\hline
$\Delta\Omega_\Lambda$&$0.0112$ &$0.0063$~($78\%$) &$0.0109$~($3\%$)   &$0.0107$~($5\%$)     &$0.0106$      &$0.0062$~($71\%$) \\\hline
$\Delta n_s$          &$0.021$  &$0.014$~($50\%$)  &$0.016$~($31\%$)   &$0.014$~($50\%$)      &$0.0112$      &$0.0096$~($17\%$) \\\hline
$\Delta A_s$          &$0.087$  &$0.053$~($64\%$)  &$0.061$~($43\% $)  &$0.049$~($77\%$)      &$0.033$       &$0.029$~($14\%$)  \\\hline
$\Delta \tau$         &$0.050$  &$0.033$~($51\%$)  &$0.038$~($32\%$)   &$0.032$~($56\%$)      &$0.026$       &$0.024$~($8\%$)   \\\hline
$\Delta b_1$          &$0.081$  &$0.051$~($59\%$)  &$0.051$~($59\%$)   &$0.036$~($125\%$)      & -            &  -                \\\hline
$\Delta b_2$          & -       &$0.041$           & -                 & -                   & -            & -                \\\hline
$\Delta \sigma_8$     &$0.064$  &$0.036$~($78\%$)  &$0.043$~($49\%$)   &$0.032$~($100\%$)      &$0.016$       &$0.011$~($45\%$)  \\\hline
$\Delta h$            &$0.0139$ &$0.0088$~($58\%$) &$0.0116$~($20\%$)  &$0.0110$~($26\%$)     &$0.0103$      &$0.0074$~($39\%$) \\\hline
$\Delta \Omega_b$     &$0.00139$&$0.00111$~($25\%$)&$0.00134$~($4\%$)  &$0.00134$~($4\%$)    &$0.00134$     &$0.00104$~($29\%$)\\\hline
\end{tabular}
\end{ruledtabular}
\end{table*}
\endgroup

As expected, the effectiveness of the bispectrum in constraining cosmology depends significantly on the smallest scale considered due to the fast rise in the number of triangles available. One can notice how, already at $k_{\smax}^{MS}=0.2\kMpc$ when combined with the power spectrum, it can improve errors by a $13$ to $56\%$. At $k_{\smax}^{MS}=0.3\kMpc$ when considered alone with CMB information the bispectrum can actually improve over the power spectrum by $6\%$ to more than $70\%$ for $\Omega_\Lambda$, although at the expense of a poorer determination of the linear bias. We should keep in mind that the bispectrum analysis introduces an extra parameter, the quadratic bias, and that one can expect a better constrain on the linear bias when combined with the power spectrum.

A quick glance at Table~\ref{tab_results_9P} shows that most of the improvement (numbers in bold) brought by the bispectrum are in parameters related to the overall amplitude of fluctuations and the effective spectral index. This is expected as the bispectrum breaks the degeneracy between bias and dark matter amplitude fluctuations~\cite{Frieman:1993nc,Fry1994}, and its configuration dependence is sensitive to the spectral index because of the anisotropy of tidal gravitational fields and velocity flows~\cite{Sco97}.

In Fig.~\ref{fig9Pk3} we compare the CMB power spectrum likelihoods to the combined power spectrum, bispectrum and LRG power spectrum likelihoods. From this and Table~\ref{tab_results_9P} one sees that most of improvement over CMB alone is coming from pairs of statistics that involve the bispectrum ({\it i.e.} either P+B or P$_L$+B, not shown for clarity). This is so because the most significant improvements arise due to breaking of degeneracies present in the LSS or CMB~\cite{EHT99}. This manifests itself in the entries in Table~\ref{tab_results_9P} ($k_{\smax}^{MS}=0.2\kMpc$) in several ways : 1) W+P$_L$ improves mildly over W+P, but W+P+P$_L$ improves significantly over W+P (consistent with Table~2 in~\cite{Eisenstein:2005su}) 2) W+P+B is better than W+P+P$_L$ in most parameters (except those related to $\Omega_m$: $\omega_d$, $\Omega_b$ and $\Omega_\Lambda$, since a better detection of the acoustic scale in the LRG sample gives a high quality constraint~\cite{Eisenstein:2005su}). This holds even though the signal to noise in B (for $k_{\smax}^{MS}=0.2\kMpc$) is not as large as in P$_L$ (e.g. compare W+B  vs. W+P$_L$), because B is more complementary than P$_L$ to W+P, {\it i.e.} using non-Gaussian information provides a substantially different direction in parameter space. When using information up to $k_{\smax}^{MS}=0.3\kMpc$, W+P+B constrains all parameters better than W+P+P$_L$ except for $\omega_d$. In this case, adding P$_L$ to W+P+B still helps in improving parameters slightly (see Fig.~\ref{fig9Pk3}), particularly for $\Omega_\Lambda$ (or $\Omega_m=1-\Omega_\Lambda$).

It is interesting to compare the results on bias parameters to those in a fixed cosmology, as assumed in past work~\cite{Scoccimarro:2000sp,Feldman:2000vk,Verde:2002ed,Gazta05}. Performing an analysis of the bispectrum alone with fixed cosmology, one finds for linear and quadratic bias the errors ($k_{\smax}^{MS}=0.3\kMpc$)
\beq
\Delta b_1 = 0.015 , \qquad \Delta b_2 = 0.045.
\eeq

Comparing this to the corresponding entry (W+B) in Table~\ref{tab_results_9P} we see that when cosmology is allowed to vary, the determination of $b_1$ suffers from the degeneracy with $A_s$ and, when combined with CMB data, with the optical depth $\tau$, while the result for $b_2$ is essentially unaffected. On the other hand, this is the price one pays for constraining cosmological parameters more accurately.

Figure~\ref{fig_9P_k3_contours} shows the marginalized 95\% CL contour plots of pairs of parameters. The role played by the bispectrum in lifting the degeneracy between the galaxy bias parameter $b_1$ and the parameter $A_s$ determining the amplitude of dark matter fluctuations is particularly evident. It is clear in particular from the $b_1$-$A_s$ contours, that the combination of power spectrum and bispectrum, by narrowing the uncertainty on these two parameters, affects the determinations of all the others. The question then arises, are the improvements on cosmological parameters brought by using the bispectrum just a result of having constrained galaxy bias?

\subsection{Not Just Galaxy Bias}
\label{NotJB}

In order to answer this question, we present in Table~\ref{results9PavgPvsB} a couple of tests that compare W+P with W+P+B for $k_{\smax}^{MS}=0.3\kMpc$. The first two columns repeat the constraints shown before in Table~\ref{tab_results_9P}, whereas the third column shows the W+P results when a prior on $b_1$ is added to mimic the W+P+B constraint on bias. Since the marginalized likelihood of $b_1$ is approximately Gaussian (see Fig.~\ref{fig9Pk3}) we can add a Gaussian prior with width $\sigma$ given by
\beq\label{b1_Gauss_prior}
\frac{1}{\sigma^2}=\frac{1}{\sigma_B^2}-\frac{1}{\sigma_P^2}
\eeq
where $\sigma_B$ is the error on $b_1$ from the W+P+B analysis and $\sigma_P$ is that from W+P. We see from Table~\ref{results9PavgPvsB} that this reproduces the W+P+B bias constraint closely enough. By comparing the rest of the entries in W+P+B against W+P+$b_1$ prior {\em it follows that the improvement on cosmological parameter determination from the bispectrum is not only due to constraining galaxy bias}. 

The right side of  Table~\ref{results9PavgPvsB} presents another test, where the bias parameters are fixed ($b_1=1$, $b_2=0$). Comparing these last two columns we see a significant improvement from adding bispectrum information. 

The fourth column in Table~\ref{results9PavgPvsB}  shows the analysis of the W+P case with a prior on linear bias, $\Delta b_1 =0.036$~\cite{Sefusatti:2004xz}, corresponding to the case where the bispectrum is analyzed through the hierarchical amplitude $Q$ [see Eq.~(\ref{Qgal})] for a fixed cosmology. We see that in this case some of the constraints agree, but the error on bias is significantly underestimated, whereas the errors on $\Omega_\Lambda$ and $h$ are significantly overestimated. Interestingly, the uncertainty in $\sigma_8$ is robust to this analysis (which is incorrect due to neglecting cross-correlations between $Q$ and $P$ and bias with cosmology).

\subsection{The Effects of Beat Coupling}
\label{likeBC}

\begin{figure}[t]
\begin{center}
\includegraphics[width=0.48\textwidth~]{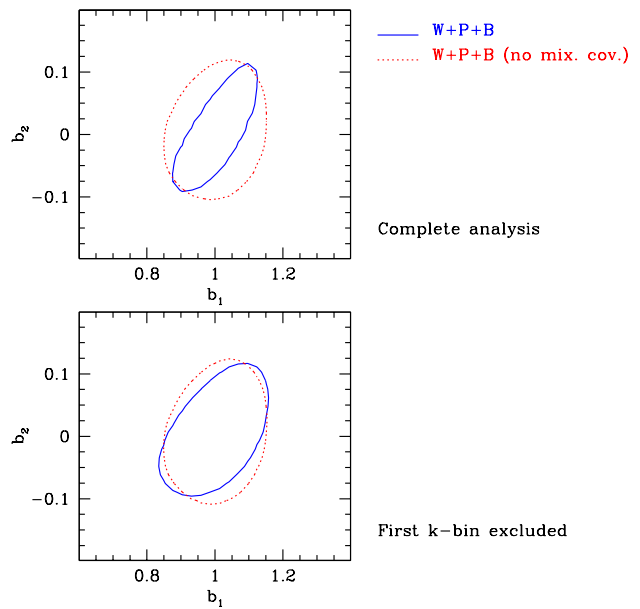}
\caption{\label{fig_9P_k3_k3r_bias_contours} Marginalized $95\%$ contour plots for the linear bias parameter $b_1$ and the quadratic bias parameter $b_2$ in the W+P+B case  with $k_{\smax}^{MS}=0.3 \kMpc$. The top panel shows the full analysis with (blue, continuous line) and without (red, dashed line) the mixed covariance matrix, whereas in the bottom panel the two analyses are repeated excluding the lowest power spectrum bin, thus suppressing beat coupling.}
\end{center}
\end{figure}

It is interesting to see what happens with the constraints on parameters if the mixed covariance between power spectrum and bispectrum is ignored, this is given in brackets in the fourth column of Table~\ref{tab_results_9P} for the W+P+B case. Naively, one would expect that excluding the mixed covariance should lead to better constraints, but as shown in Table~\ref{tab_results_9P} this is incorrect for most parameters: this is due to the effects of beat coupling.

As discussed in section~\ref{secPBcov}, beat coupling means that the structure of the mixed covariance matrix is dominated by up and down correlated fluctuations of the whole power spectrum and bispectrum of narrow isosceles triangles  depending on the power of the largest mode in the survey, as shown in Eq.~(\ref{beatCeq25}). Not allowing for such effect in the covariance matrix means that these fluctuations will be mistaken as a signature of larger errors in the parameters that characterize the amplitude of galaxy fluctuations since these are the parameters that can mimic such behaviour. Indeed, as seen by comparing the third and fourth columns in Table~\ref{tab_results_9P}, including the mixed covariance (thus allowing for beat coupling) reduces the errors mostly on $A_s$, $\tau$, $b_1,b_2$ and thus $\sigma_8$.

In Fig.~\ref{fig_9P_k3_k3r_bias_contours} we illustrate this point further by showing the marginalized $95\%$ contour plots for the linear bias parameter $b_1$ and the quadratic bias parameter $b_2$ in the W+P+B case  with $k_{\smax}^{MS}=0.3 \kMpc$. In the top panel we show the full analysis, whereas the bottom panel drops the bin with the lowest value of $k$ in the power spectrum. The top panel shows a significant difference between including the mixed terms $\lan \d P_i \d B_j \ran$ in the complete covariance matrix (solid) and dropping them (dashed). We see that including the mixed covariance gives clearly a tighter constraint on the two parameters together with a slight degeneracy. 

\begin{figure}[t!]
\begin{center}
\includegraphics[width=0.48\textwidth~]{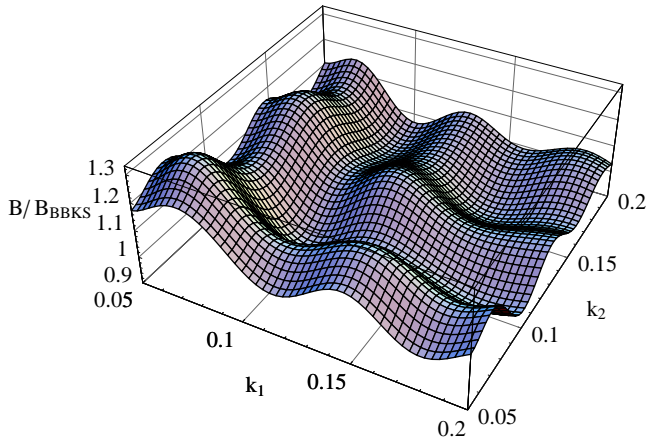}
\caption{\label{fig_BispWiggles} Baryon acoustic oscillations imprinted in the bispectrum as a function of $k_1$ and $k_2$ for fixed angle given by $\hat{k}_1 \cdot \hat{k}_2=-1/2$. The diagonal $k_1=k_2$ corresponds to equilateral triangles. As the angle between the two vectors is varied the pattern of peaks moves accordingly. }
\end{center}
\end{figure}

\begin{figure*}[t]
\begin{center}
\includegraphics[width=0.98\textwidth~]{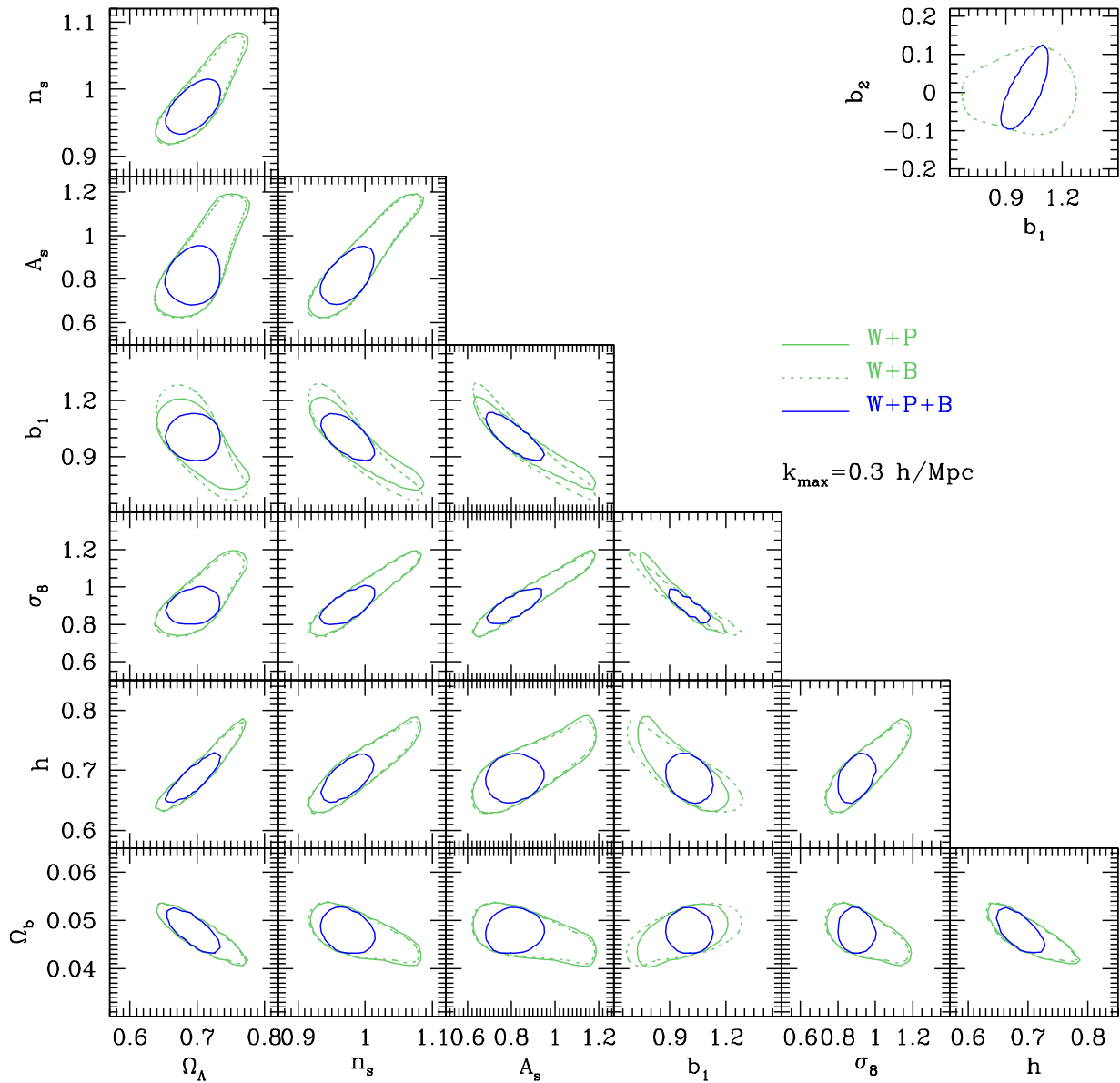}
\caption{\label{fig_9P_k3_contours_bbks} Same as Fig.~\ref{fig_9P_k3_contours} but for a featureless (no baryon acoustic oscillations)  transfer function.}
\end{center}
\end{figure*}
In the lower panel the same contours are plotted but now the analysis excludes the power spectrum bin corresponding to the smallest $k$-value, thus suppressing the effects of beat coupling. We see that in this case there is not much difference between including or not the mixed covariance, in fact the mild degeneracy induced by including the mixed covariance leads to slightly larger errors for $b_1$ and $b_2$. The same behavior is seen with all other parameters when the first $k$-bin is excluded, except for $A_s$ and $\tau$ which still show a minor improvement when the mixed covariance is included. This is likely due to residual beat coupling, {\it e.g.} a careful look at the left panel in Fig.~\ref{densityPB} shows that vertical features also exist for modes with $k=2\Delta k$, although at a much lower amplitude.

\begin{table*}[t!]
\caption{\label{tab_results_10P} Same as Table~\ref{tab_results_9P} but for $w$CDM models}
\begin{ruledtabular}
\begin{tabular}{l|lllll|ll}
  & W+P & W+B & W+P+B & W+P+B (no mix. cov.) & W+P+B+P$_L$ & W+P+P$_L$ & W+P$_L$ \\
\hline\hline
 & \multicolumn{5}{l}{$k_{\smax}^{MS}=0.2 \kMpc$} & \multicolumn{2}{l}{$k_{\smax}^{LRG}=0.2 \kMpc$}\\\hline\hline
$\Delta\omega_d$      &$0.0070$ &$0.0073$~($-4\%$) &$0.0061$~($15\%$) &[~$0.0058$~($21\%$)~] &$0.0047$~($49\%$) &$0.0050$~($40\%$) &$0.0056$~($25\%$)\\\hline
$\Delta\omega_b$      &$0.00109$&$0.00113$~($-4\%$)&$0.00094$~($16\%$)&[~$0.00097$~($12\%$)~]&$0.00084$~($30\%$)&$0.00094$~($16\%$)&$0.00102$~($7\%$)\\\hline
$\Delta\Omega_\Lambda$&$0.0141$ &$0.0121$~($17\%$) &$0.0102$~($38\%$) &[~$0.0098$~($44\%$)~] &${\bf 0.0076}$~($86\%$) &${\bf 0.0089}$~($58\%$) &$0.0101$~($40\%$)\\\hline
$\Delta n_s$          &$0.032$  &$0.032$~($0\%$)   &$0.026$~($23\%$)  &[~$0.027$~($19\%$)~]  &$0.022$~($45\%$)  &$0.025$~($28\%$)  &$0.029$~($10\%$)  \\\hline
$\Delta A_s$            &$0.131$  &$0.129$~($2\%$)   &$0.107$~($22\%$)  &[~$0.112$~($17\%$)~]  &$0.092$~($42\%$)  &$0.109$~($20\%$)  &$0.119$~($10\%$)  \\\hline
$\Delta w$            &$0.107$  &$0.103$~($4\%$)   &$0.098$~($9\%$)   &[~$0.096$~($11\%$)~]  &$0.084$~($27\%$)  &$0.089$~($20\%$)  &$0.095$~($13\%$) \\\hline
$\Delta \tau$         &$0.077$  &$0.075$~($3\%$)   &$0.067$~($15\%$)  &[~$0.068$~($13\%$)~]  &$0.058$~($33\%$)  &$0.066$~($17\%$)  &$0.071$~($8\%$)  \\\hline
$\Delta b_1$          &$0.099$  &$0.132$~($-25\%$) &${\bf 0.064}$~($55\%$)  &[~$0.079$~($25\%$)~]  &${\bf 0.055}$~($80\%$)  &$0.073$~($36\%$)  & -               \\\hline
$\Delta b_2$          & -       &$0.072$           &$0.055$           &[~$0.063$~]           &$0.051$           & -                & -               \\\hline
$\Delta b_1^{LRG}$    & -       & -                & -                & -                    &$0.101$           &$0.129$           &$0.14$           \\\hline\hline
$\Delta \sigma_8$     &$0.076$  &$0.085$~($-11\%$) &${\bf 0.049}$~($55\%$)  &[~$0.057$~($33\%$)~]  &${\bf 0.042}$~($81\%$)  &$0.056$~($36\%$) &$0.066$~($15\%$) \\\hline
$\Delta h$            &$0.0183$ &$0.0164$~($12\%$) &${\bf 0.0115}$~($59\%$) &[~$0.0123$~($49\%$)~] &${\bf 0.0096}$~($91\%$) &${\bf 0.0119}$~($54\%$) &$0.014$~($31\%$) \\\hline
$\Delta \Omega_b$     &$0.0028$ &$0.0025$~($12\%$) &$0.0022$~($27\%$) &[~$0.0022$~($27\%$)~] &${\bf 0.0017}$~($65\%$) &$0.0019$~($47\%$) &$0.0022$~($27\%$)\\\hline
\hline
 & \multicolumn{5}{l}{$k_{\smax}^{MS}=0.3 \kMpc$} & \multicolumn{2}{l}{$k_{\smax}^{LRG}=0.2 \kMpc$}\\\hline\hline
$\Delta\omega_d$      &$0.0068$ &$0.0056$~($21\%$) &$0.0054$~($26\%$) &[~$0.0053$~($28\%$)~] &${\bf 0.0044}$~($54\%$) &$0.0050$~($36\%$) & \\\hline
$\Delta\omega_b$      &$0.00107$&$0.00094$~($14\%$)&$0.00087$~($23\%$)&[~$0.00083$~($29\%$)~]&$0.00080$~($34\%$)&$0.00092$~($16\%$)& \\\hline
$\Delta\Omega_\Lambda$&$0.0120$ &$0.0083$~($45\%$) &$0.0082$~($46\%$) &[~$0.0074$~($62\%$)~] &${\bf 0.0066}$~($82\%$) &$0.0084$~($43\%$) & \\\hline
$\Delta n_s$          &$0.031$  &$0.026$~($19\%$)  &$0.024$~($29\%$)  &[~$0.022$~($41\%$)~]  &$0.021$~($48\%$)  &$0.025$~($24\%$)  & \\\hline
$\Delta A_s$            &$0.129$  &$0.110$~($17\%$)  &$0.096$~($34\%$)  &[~$0.094$~($37\%$)~]  &${\bf 0.084}$~($54\%$)  &$0.108$~($19\%$)  & \\\hline
$\Delta w$            &$0.105$  &$0.090$~($17\%$)  &$0.091$~($15\%$)  &[~$0.085$~($23\%$)~]  &$0.080$~($31\%$)  &$0.088$~($19\%$)  & \\\hline
$\Delta \tau$         &$0.076$  &$0.066$~($15\%$)  &$0.061$~($25\%$)  &[~$0.059$~($29\%$)~]  &$0.053$~($43\%$)  &$0.065$~($17\%$)  & \\\hline
$\Delta b_1$          &$0.091$  &$0.100$~($-9\%$) &${\bf 0.052}$~($75\%$)  &[~$0.062$~($47\%$)~]  &${\bf 0.047}$~($94\%$)  &$0.071$~($28\%$)  & \\\hline
$\Delta b_2$          & -       &$0.046$           &$0.042$           &[~$0.045$~]           &$0.040$           & -                & \\\hline
$\Delta b_1^{LRG}$    & -       & -                & -                & -                    &$0.085$           &$0.126$           & \\\hline\hline
$\Delta \sigma_8$     &$0.070$  &$0.062$~($13\%$)  &${\bf 0.038}$~($84\%$)  &[~$0.042$~($67\%$)~]  &${\bf 0.035}$~($100\%$)  &$0.055$~($27\%$)  & \\\hline
$\Delta h$            &$0.0160$ &$0.0108$~($48\%$) &${\bf 0.0091}$~($76\%$) &[~$0.0089$~($80\%$)~] &${\bf 0.0083}$~($93\%$) &$0.0114$~($40\%$) & \\\hline
$\Delta \Omega_b$     &$0.0025$ &$0.0018$~($39\%$) &$0.0018$~($39\%$) &[~$0.0017$~($47\%$)~] &${\bf 0.0016}$~($56\%$) &$0.0019$~($32\%$) & \\\hline
\end{tabular}
\end{ruledtabular}
\end{table*}

\subsection{Baryon Acoustic Oscillations}
\label{secBAO}

The same baryon acoustic oscillation features induced in the dark matter power spectrum~\cite{PeYu70,BoEf84} and recently seen in galaxy surveys~\cite{Eisenstein:2005su, Cole2005} are expected to be present in the bispectrum~\cite{Josh} and can also be used to help in determining cosmological parameters. Figure~\ref{fig_BispWiggles} shows the ratio of the bispectrum to a featureless (no acoustic oscillation) bispectrum obtained from the BBKS fitting formula~\cite{Bardeen:1985tr} by setting the shape parameter $\Gamma=0.175$. Because the bispectrum scales as the square of the power spectrum, the $15\%$ modulation in power leads to a $30\%$ modulation in the bispectrum. At $k=0.1 \kMpc$ the signal to noise in the bispectrum is about twice smaller than for the power spectrum~\cite{Sefusatti:2004xz}, so this scale roughly presents the limit after which the bispectrum gives a better constraint on acoustic oscillations than the power spectrum. Unfortunately, at $z \simeq 0$ the acoustic oscillations are washed out by nonlinearities for $k \ga 0.1 \kMpc$~\cite{MWP99, SeEi05, Springel05, White05}.

A fair assessment of the improvement on the detection of acoustic oscillations by using the bispectrum is beyond the scope of this paper and will be presented elsewhere. Here we note that our mock catalogs, although not exact in their nonlinear properties, do include the suppression of acoustic features, that is, we do not assume Eulerian second-order perturbation theory as done in~Fig.~\ref{fig_BispWiggles} for illustrative purposes.

 In order to assess the impact of acoustic features in our study we compute marginalized likelihoods using instead the BBKS fitting formula for the transfer function, the results are shown in Fig.~\ref{fig_9P_k3_contours_bbks} where we reproduce the same marginalized 95\% CL contour plots given in Fig.~\ref{fig_9P_k3_contours}. Since this transfer function depends exclusively on the shape parameter $\Gamma=\Omega_m h \exp[-\Omega_b(1+\sqrt{2h}/\Omega_m)]$~\cite{Sugi95}, we generically expect an enhanced degeneracy between $\Omega_m$ (or, equivalently here, $\Omega_\Lambda$) and the Hubble parameter $h$.  One can immediately notice in general a stronger degeneracy in all contour plots and, in particular, a more similar behavior of the bispectrum and power spectrum contours, especially for those involving $\Omega_\Lambda$. This is the degeneracy that gets broken by acoustic features, as it is well known in the power spectrum case~\cite{EHT99}. 

\begin{figure*}[t]
\begin{center}
\includegraphics[width=0.98\textwidth~]{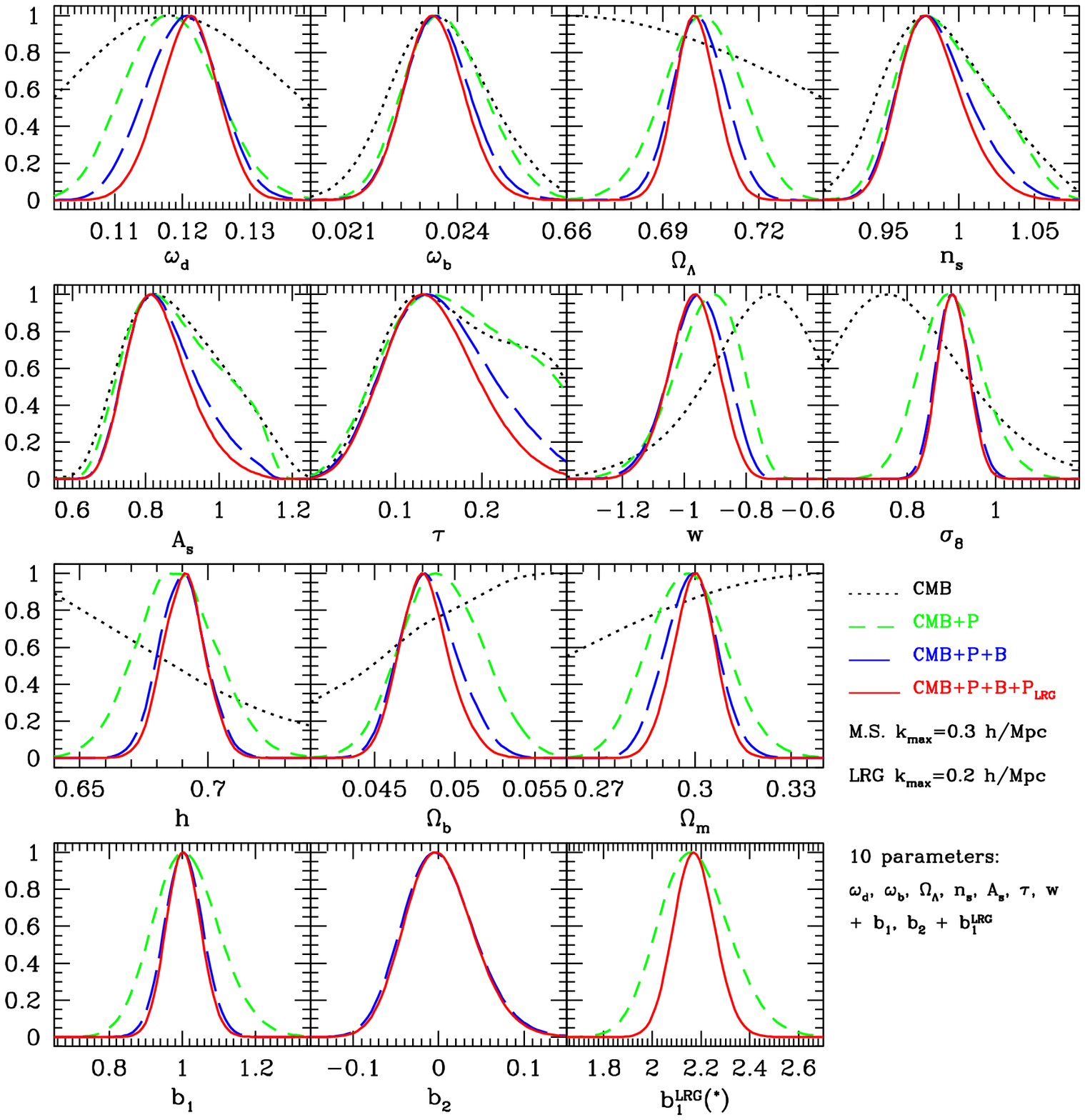}
\caption{\label{10PkM0p30single} $w$CDM models: marginalized likelihood functions for the six+four cosmological parameters assuming $k_{\smax}^{MS}=0.3 \kMpc$. For the $b_1^{LRG}$ parameter, the dotted line denotes the WMAP plus LRG power spectrum likelihood.}
\end{center}
\end{figure*}

Focusing for instance on the $\Omega_\Lambda$ {\it vs.} $h$ case in Fig.~\ref{fig_9P_k3_contours_bbks} and Fig.~\ref{fig_9P_k3_contours}, we can see that the bispectrum, by virtue of its several different triangular configurations, is remarkably sensitive to features in the dark matter linear power spectrum such as the baryonic acoustic oscillations.  The marginalized errors on individual parameters are, overall, larger when the BBKS transfer function is used. We notice, however, that the improvement provided by adding the bispectrum improves for parameters such as $A_s$, $\tau$, $b_1$ and the spectral index $n_s$ (about a factor of two better than the power spectrum alone) while it reduces for $\Omega_\Lambda$.

\subsection{Dark Energy: $w$CDM models} 
\label{secwCDM}

\begin{figure*}[t]
\begin{center}
\includegraphics[width=0.98\textwidth~]{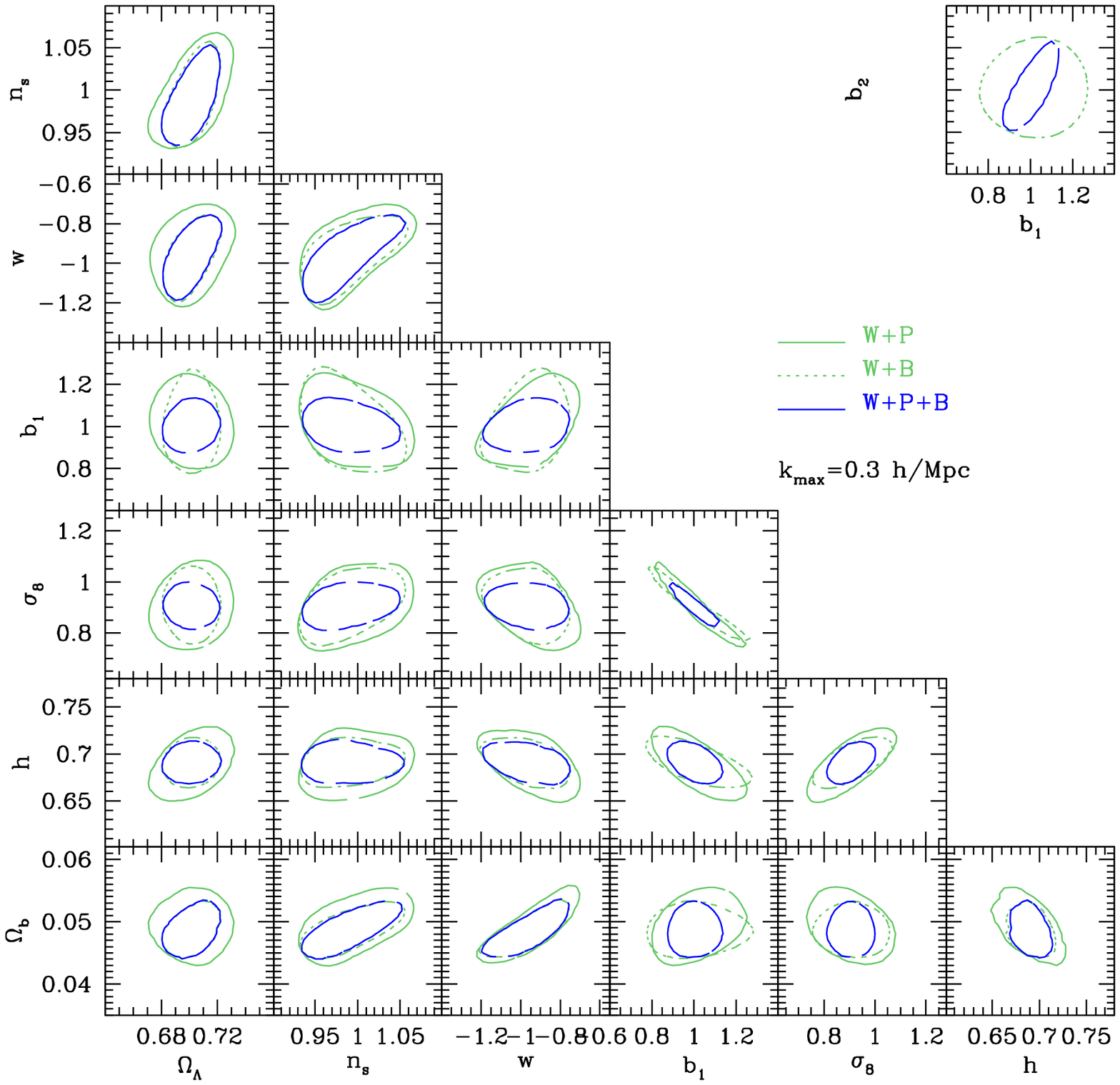}
\caption{$w$CDM models: marginalized $95\%$ contour plots for a selection of the cosmological parameters assuming $k_{\smax}^{MS}=0.3 \kMpc$.}\label{10Pk3contours}
\end{center}
\end{figure*}

We now extend the analysis performed above to include the determination of the dark energy equation of state parameter $w$ under the assumption of an homogeneous dark energy component. We assume $w$ to be constant. 
Introducing $w$, on which the growth function depends, leads to an increased degeneracy with the other parameters controlling the amplitude of the galaxy fluctuations such as $A_s$ and $b_1$, which can be ameliorated by including bispectrum information. 

Table~\ref{tab_results_10P} presents the expected errors on the various parameters for the two cases of $k_{\smax}^{MS}=0.2\kMpc$ and $0.3\kMpc$. We find that for  $k_{\smax}^{MS}=0.2\kMpc$ the determination of the $w$ parameter improves, over the W+P case, by 10\% when the bispectrum is included, while by comparison a  20\% if the LRG power spectrum is added instead. For  $k_{\smax}^{MS}=0.3\kMpc$ adding the bispectrum improves the determination of $w$ by 15\%. These are mild improvements, but note that there are basically three different ways of getting below 10\% errors on $w$ by using the power spectrum of main and LRG samples, and the non-Gaussian information in the bispectrum and thus important consistency checks. Of course, the addition of extra information such as type IA supernovae or weak gravitational lensing (apart from the latest CMB data) will tighten the constraints.

In Fig.~\ref{10PkM0p30single} we plot the marginalized likelihood functions for the case where the maximum wavenumber is $k_{\smax}^{MS}=0.3 \kMpc$. Note that unlike the $\Lambda$CDM models previously considered, some of the maximum likelihood values of the first year WMAP data differ substantially from the chosen fiducial values for the LSS likelihood function. For instance, WMAP  gives a marginalized probability distribution for $w$ with a maximum close to $-0.72$, rather far from our fiducial value $w=-1$. This implies that part of the constraing power of the LSS statistics is spent in shifting the maximum to the  $w=-1$. 

Finally, in Fig.~\ref{10Pk3contours} we show the contour plots for some selection of parameters. Comparing these results to the previous on $\Lambda$CDM models we see that the improvement brought by the bispectrum is increased in the case of $b_1$ and $\sigma_8$, and in fact the final error bars on these parameters (and $b_2$) are almost insensitive to including a more generic dark energy. This is good news as $\sigma_8$ is one of the least known parameters and subject to tension between different data sets~\cite{Spergel:2006hy,Brown03,Hoekstra05,VHL06,SSM06}.

On the other hand, the constraints on $\Omega_\Lambda$, $A_s$, $\tau$ and particularly $\omega_d$ and $n_s$ are significantly worse than in the cosmological constant case. Regarding the behavior of the mixed covariance matrix, we see the same impact of beat coupling that we discussed before, the parameters responsible for the amplitude of galaxy fluctuations $A_s$, $\tau$, $\sigma_8$ and bias parameters improve by the inclusion of the mixed covariance, though the behavior of the rest of the parameters is somewhat more complicated. Performing the same test as in Fig.~\ref{fig_9P_k3_k3r_bias_contours} in this case we see that excluding the first $k$-bin completely erases the effects due to beat coupling.

\section{Conclusions}
\label{secConclusions}

We have provided a first detailed study on the information about cosmological parameters contained in the bispectrum of the galaxy distribution at large scales, paying particular attention to the joint analysis with the power spectrum and their combination with CMB data. We have shown that the bispectrum has significant information on cosmological parameters and when combined with the power spectrum it is more complementary than combining power spectra of different samples of galaxies, since non-Gaussianity provides a somewhat different direction in parameter space.  Moreover, replacing the power spectrum with the bispectrum gives similar constraints on cosmological parameters and can therefore serve as a consistency check.

In order to properly combine bispectrum with power spectrum information, we worked out their covariance properties. Due to the effects of beat coupling~\cite{Hamilton:2005dx}, the mixed terms in the covariance matrix are enhanced. We demonstrated  that including this effect into the likelihood analysis provides a slight improvement on the error bars of cosmological parameters related to the amplitude of galaxy fluctuations.

In the framework of flat cosmological models we showed that most of the improvement of adding bispectrum information corresponds to parameters related to the amplitude and effective spectral index of perturbations, in particular $\Omega_m$ (or $\Omega_\Lambda=1-\Omega_m$) and $\sigma_8$, which can be improved by factors of 1.5 to 2 and, interestingly, are presently among the least well known. In particular, we showed that the uncertainties on $\sigma_8$ are robust to relaxing the equation of state parameter $w$ beyond a cosmological constant. This is good news as $\sigma_8$ is subject to tension between different data sets~\cite{Spergel:2006hy,Brown03,Hoekstra05,VHL06,SSM06}. We also showed that the improvements are not directly a consequence of just constraining galaxy bias but of genuine information on cosmological parameters.

As far as future theoretical work is concerned, the single most pressing issue is that of systematic errors in the predictions, which were addressed in~\cite{Sco00b} for a previous generation of galaxy surveys. Those methods, based on second-order Lagrangian perturbation theory (which we have used here), are likely not enough given the expected statistical errors we derived in this work. Fortunately, powerful methods based on first principles have recently become available~\cite{RPT}, and together with numerical simulations they should provide a sound theoretical basis. Work on this is in progress and will be reported soon.

\vspace{0.8cm}

\acknowledgments

We thank the referee for a careful reading of the manuscript and comments that improve the presentation of this paper. 
We thank Josh Frieman for helpful comments and Mulin Ding for his technical assistance and his remarkable patience in maintaining our aging computer cluster Mafalda. We wish to thank Alan Sokal for giving us access to his group's computer cluster Guille, which was funded in part by NSF grant PHY--0424082. We thank NYU Information Technology Services for making  its High Performance Computation Cluster available to us, and thank Joseph Hargitai for his technical assistance. E.~S. is supported by the US Department of Energy and by NASA grant NAG5-10842 at Fermilab.

\appendix

\section*{Appendix: WMAP 3-year update}
\label{secResults_3y}

Shortly after the completion of the present work, the 3-year WMAP satellite observations, \cite{Spergel:2006hy}, became publicly available. For the reasons described below, the recent data is not simply an incremental improvement on the results presented in the main sections of the paper, and it is instructive to see the differences in the constraining power of the bispectrum. For this reason, and in order not to introduce substantial changes to preprint version of the paper, we add this appendix updating the results of section~\ref{secResults} to include the CMB likelihood corresponding to the new data. 

The improved analysis of the CMB polarization on the 3-year data is responsible for significant differences with respect to the 1-year results in the errors on cosmological parameters as well as in their best values. Particularly significant to the present work is the much improved determination of the optical depth parameter $\tau$ leading to a better constraints on the amplitude of primordial fluctuations $A_s$ and on the spectral index $n_s$. 

We can generically expect that the smaller errors from the CMB analysis alone would make, for certain parameters, the effect of adding other data sets less noticeable. In particular we can expect a reduced relative impact of including the bispectrum in the large-scale structure data analysis with respect to the parameters responsible for the amplitude of galaxy fluctuations. 

However, as we will show, the constraining power of the power spectrum and bispectrum joint analysis turns to other relevant parameters, particularly in the case of the $w$CDM models, where the uncertainty on the dark energy equation of state introduces substantial degeneracies while the large-scale structure statistics are particularly sensitive to the late-time expansion via the growth factor $D(\pv;a)$.

Another reason to expect the improvement due to the bispectrum likelihood to be somehow smaller is related to the lower value of the amplitude parameter $A_s$ and, consequently, of $\sigma_8$. From Eq.~(\ref{BispEPT}) and Eq.~(\ref{covBS}) one can derive the signal to noise for a given triangular configuration, assumed here equilateral for simplicity, so that, at large scales, we have
\beq
\left(\frac{S}{N}\right)(k)\equiv\frac{B(k,k,k)}{\Delta B(k,k,k)}\sim \sqrt{P(k)}\sim\sqrt{A_s}.
\eeq 
The best value for the parameter $A_s$ decreased from the 1-year to 3-year WMAP analysis by almost $15\%$. On the other hand a lower value for amplitude of primordial fluctuations results as well in a lower value for the non-linear scale thereby making more robust our approximations based on perturbation theory.   

\begin{table}[t]
\caption{\label{fiducialvalues6P_3y} Fiducial values for the cosmological and bias parameters assumed for the LSS+WMAP 3-year likelihood analysis.}
\begin{ruledtabular}
\begin{tabular}{llcc}
Parameter                         &  $\Lambda$CDM models & $w$CDM models \\\hline
$\omega_d$              & $0.104$   & $0.105$     \\
$\omega_b$                   & $0.0223$  & $0.0221$      \\
$\Omega_\Lambda$                 & $0.765$   & $0.715$    \\
$n_s$                        & $0.95$    & $0.944$     \\
$A_s$                   & $0.687$ & $0.715$         \\
$w$                    & $-1$      & $-0.84 $      \\
$\tau$                    & $0.09$    & $0.09$     \\
$b_1$                & $1$       & $1$      \\
$b_2$              & $0$       & $0$      \\
$b_1^{LRG}$                  & $2.17$    & $2.17$      \\
\end{tabular}
\end{ruledtabular}
\end{table}

\begingroup
\begin{table*}[t!]
\caption{\label{tab_results_9P_3y} $\Lambda$CDM models: expected marginalized errors ($68\%$ CL) for WMAP3 (temperature and polarization, W) combined with the SDSS main sample power spectrum (P) and bispectrum (B) and with the LRG power spectrum (P$_L$). The percentage in parenthesis indicates the improvement over the analysis including the main sample power spectrum alone (W+P), numbers in bold indicate errors down by at least $1.5$. In brackets we quote the W+P+B errors obtained by ignoring the mixed power spectrum - bispectrum covariance. Compare with Table~\ref{tab_results_9P} for WMAP1.}
\begin{ruledtabular}
\begin{tabular}{l|lllll|ll}
  & W+P & W+B & W+P+B & W+P+B (no mix. cov.) & W+P+B+P$_L$ & W+P+P$_L$ & W+P$_L$ \\
\hline\hline
 & \multicolumn{5}{l}{$k_{\smax}^{MS}=0.2 \kMpc$} & \multicolumn{2}{l}{$k_{\smax}^{LRG}=0.2 \kMpc$}\\\hline\hline
$\Delta\omega_d$      &$0.0024$ &$0.0025$~($-4\%$) &$0.0020$~($20\%$) &[~$0.0020$~($20\%$)~] &$0.0017$~($41\%$)      &$0.0018$~($33\%$)      &$0.0020$~($20\%$)  \\\hline
$\Delta\omega_b$      &$0.00062$&$0.00063$~($-2\%$)&$0.00058$~($7\%$) &[~$0.00059$~($5\%$)~] &$0.00056$~($11\%$)     &$0.00058$~($7\%$)      &$0.00060$~($3\%$)  \\\hline
$\Delta\Omega_\Lambda$&$0.0088$ &$0.0077$~($14\%$) &$0.0061$~($44\%$) &[~$0.0060$~($47\%$)~] &${\bf 0.0045}$~($96\%$)&${\bf 0.0052}$~($69\%$)&$0.0062$~($42\%$)  \\\hline
$\Delta n_s$          &$0.0132$ &$0.0135$~($-2\%$) &$0.0120$~($10\%$) &[~$0.0123$~($7\%$)~]  &$0.0108$~($22\%$)      &$0.0116$~($14\%$)      &$0.0123$~($7\%$)    \\\hline
$\Delta A_s$          &$0.038$  &$0.039$~($-3\%$)  &$0.034$~($12\%$)  &[~$0.036$~($6\%$)~]   &$0.034$~($12\%$)       &$0.037$~($3\%$)        &$0.037$~($3\%$)    \\\hline
$\Delta \tau$         &$0.025$  &$0.025$~($0\%$)   &$0.023$~($8\%$)   &[~$0.024$~($4\%$)~]   &$0.023$~($8\%$)        &$0.025$~($0\%$)        &$0.049$~($6\%$)    \\\hline
$\Delta b_1$          &$0.047$  &$0.059$~($-20\%$) &$0.041$~($15\%$)  &[~$0.046$~($2\%$)~]   &$0.038$~($24\%$)       &$0.043$~($9\%$)        & -                 \\\hline
$\Delta b_2$          & -       &$0.067$           &$0.000$           &[~$0.060$~]           &$0.044$                & -                     & -                 \\\hline
$\Delta b_1^{LRG}$    & -       & -                & -                & -                    &$0.074$                &$0.081$                &$0.137$            \\\hline
\hline
$\Delta \sigma_8$     &$0.030$  &$0.031$~($-3\%$)  &$0.026$~($15\%$)  &[~$0.028$~($7\%$)~]   &$0.024$~($25\%$)       &$0.027$~($11\%$)       &$0.029$~($3\%$)   \\\hline
$\Delta h$            &$0.0112$ &$0.0098$~($14\%$) &$0.0086$~($30\%$) &[~$0.0087$~($29\%$)~] &${\bf 0.0073}$~($53\%$)&$0.0082$~($37\%$)      &$0.0089$~($26\%$)  \\\hline
$\Delta \Omega_b$     &$0.00102$&$0.00095$~($7\%$) &$0.00080$~($27\%$)&[~$0.00079$~($29\%$)~]&$0.00068$~($50\%$)     &$0.00074$~($38\%$)     &$0.00082$~($24\%$) \\\hline
\hline
 & \multicolumn{5}{l}{$k_{\smax}^{MS}=0.3 \kMpc$} & \multicolumn{2}{l}{$k_{\smax}^{LRG}=0.2 \kMpc$}\\\hline\hline
$\Delta\omega_d$      &$0.0023$ &$0.0020$~($15\%$)      &$0.0019$~($21\%$)      &[~$0.0018$~($28\%$)~] &$0.0017$~($35\%$)       &$0.0018$~($22\%$)  & \\\hline
$\Delta\omega_b$      &$0.00061$&$0.00059$~($3\%$)      &$0.00056$~($9\%$)      &[~$0.00055$~($11\%$)~]&$0.00054$~($13\%$)      &$0.00057$~($7\%$) & \\\hline
$\Delta\Omega_\Lambda$&$0.0080$ &${\bf 0.0049}$~($63\%$)&${\bf 0.0049}$~($63\%$)&[~$0.0043$~($86\%$)~] &${\bf 0.0039}$~($105\%$)&${\bf 0.0051}$~($57\%$)  & \\\hline
$\Delta n_s$          &$0.0128$ &$0.0121$~($6\%$)       &$0.0111$~($15\%$)      &[~$0.0105$~($22\%$)~] &$0.0102$~($25\%$)       &$0.0113$~($13\%$)   & \\\hline
$\Delta A_s$          &$0.038$  &$0.037$~($3\%$)        &$0.031$~($22\%$)       &[~$0.034$~($12\%$)~]  &$0.031$~($22\%$)        &$0.036$~($6\%$)   & \\\hline
$\Delta \tau$         &$0.025$  &$0.025$~($0\%$)        &$0.022$~($14\%$)       &[~$0.024$~($4\%$)~]   &$0.022$~($14\%$)        &$0.025$~($0\%$)    & \\\hline
$\Delta b_1$          &$0.046$  &$0.056$~($-18\%$)      &$0.038$~($21\%$)       &[~$0.042$~($9\%$)~]   &$0.036$~($28\%$)        &$0.042$~($9\%$)   & \\\hline
$\Delta b_2$          & -       &$0.045$                &$0.036$                &[~$0.044$~]           &$0.035$                 & -                  & \\\hline
$\Delta b_1^{LRG}$    & -       & -                     & -                     & -                    &$0.068$                 & $0.080$            & \\\hline\hline
$\Delta \sigma_8$     &$0.029$  &$0.029$~($0\%$)        &$0.023$~($26\%$)       &[~$0.024$~($21\%$)~]  &$0.022$~($32\%$)        &$0.027$~($7\%$)   & \\\hline
$\Delta h$            &$0.0103$ &$0.0076$~($35\%$)      &$0.0071$~($45\%$)      &[~$0.0080$~($65\%$)~] &${\bf 0.0065}$~($58\%$) &$0.0079$~($30\%$)  & \\\hline
$\Delta \Omega_b$     &$0.00096$&$0.00073$~($31\%$)     &$0.00073$~($31\%$)     &[~$0.00068$~($41\%$)~]&$0.00067$~($43\%$)      &$0.00074$~($30\%$) & \\
\end{tabular}
\end{ruledtabular}
\end{table*}

The definitions of the power spectrum and bispectrum likelihood functions, including their covariance properties, are the same as the ones described in the previous sections. The sole exception that we will consider for this appendix regards the fiducial values assumed for the large-scale structure observables. To make the comparison between power spectrum analysis and joint analysis as clear as possible, we take them to coincide with the maximum values of the WMAP 3-year likelihood. They are therefore different for $\Lambda$CDM and $w$CDM models and are reported in Table~\ref{fiducialvalues6P_3y}. We make use of polynomial fits to the CMB likelihood functions determined in the same fashion as explained in section~\ref{subsectionCMB} but, this time, making use of the Monte Carlo Markov chains publicly available instead of evaluating the likelihood function directly.

\subsection{$\Lambda$CDM models}
\label{secLCDM_3y}

We present in Table~\ref{tab_results_9P_3y} the expected marginalized errors (1-$\sigma$) from the power spectrum and bispectrum likelihood analysis combined with the WMAP 3-year data. As in section~\ref{secResults}, numbers in bold correspond to improvements larger than $50\%$, with the improvement factor defined in Eq.~(\ref{improve}).

General considerations such as the dependence on the smallest scale (largest wavenumber $k_{\smax}$) included are substantially unchanged. One can notice how in the $k_{\smax}=0.2\kMpc$ case the improvement due to the bispectrum ranges from $7\%$ for $\omega_b$ to $44\%$ for $\Omega_\Lambda$, while in the $k_{\smax}=0.3\kMpc$ case we have $9\%$ to $63\%$. These results are just slightly lower than those obtained with the 1-year WMAP data. Instead, the improvements on the parameters responsible for the overall amplitude of galaxy fluctuations and the spectral index are significantly reduced with respect to the analysis with WMAP 1-year data, this being a consequence, as mentioned above, of the much better determination of the parameters $\tau$ and $A_s$ from the CMB data alone, whose error bars shrinked by a factor of $2$ or more. This results, however, in much smaller expected errors on the bias parameters, with the linear bias $b_1$ determined to better than $4\%$. We can still conclude, that the combination of CMB observations with the main sample power spectrum and bispectrum obtains better constraints than the combination with power spectra from the two different samples considered in this work.

\begin{figure*}[t]
\begin{center}
\includegraphics[width=0.98\textwidth~]{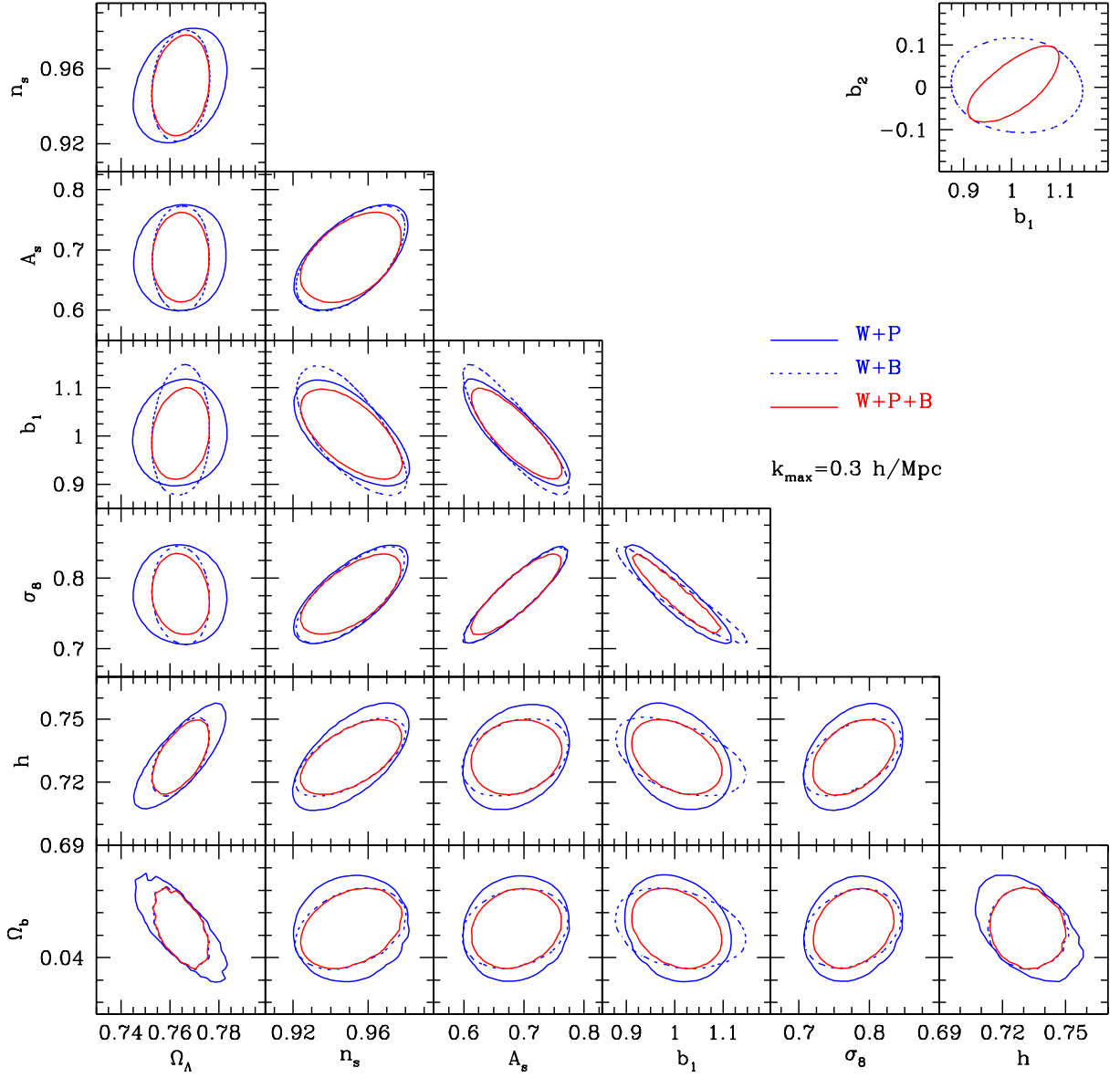}
\caption{\label{fig_9P_k3_contours_3y} Marginalized $95\%$ contour plots for pairs of cosmological parameters in the $\Lambda$CDM models assuming $k_{\smax}^{MS}=0.3 \kMpc$ and WMAP3 data. Compare with Fig.~\ref{fig_9P_k3_contours} for WMAP1.}
\end{center}
\end{figure*}

Such considerations are reflected in the marginalized 95\% CL contour plots shown in Figure~\ref{fig_9P_k3_contours_3y}, where we can still observe the different directions of several degeneracies of the power spectrum likelihood function with respect to the bispectrum one. 

We have checked that the conclusions derived from Table~\ref{results9PavgPvsB} and discussed in section~\ref{NotJB} hold as well for the WMAP 3-year data.  In both tests, we can still observe better results in the W+P+B analysis, particularly for certain cosmological parameters such as $\Omega_m$, again showing that the information provided by the bispectrum is not limited to the galaxy bias determination.

\subsection{Dark Energy: $w$CDM models} 
\label{secwCDM_3yr}

Finally, in Table~\ref{tab_results_10P_3y} we present the expected marginalized errors (1-$\sigma$) on the cosmological parameters for models which allow for a homogeneous dark energy component with an equation of state parametrized by the constant $w$.

As already mentioned, for these models the improvement due to the new CMB data set to the constraints on the parameters responsible for the amplitude of galaxy fluctuations is now less dramatic because of the degeneracy with $w$. In this case, therefore, we can expect a more relevant contribution of the bispectrum to the combined analysis because of the dependence of large-scale structure statistics on the growth function $D(\pv,a)$. 

We see indeed that, for $k_{\smax}=0.3\kMpc$, including the bispectrum, the error on $w$ improves by $46\%$, a significantly larger improvement compared to the previous analysis using WMAP1 ($15\%$), or compared to the results obtained by adding the LRG power spectrum to the main sample one and WMAP3 ($24\%$). We also find that in the most general case of W+P+B+P$_L$ the same error gets better by a $54\%$. Similar observations hold as well for other parameters such as $\omega_d$ and $\Omega_\Lambda$ while we notice less impressive results for $A_s$ and $n_s$. The corresponding two-parameter marginalized contour plots for the $w$CDM models updated to WMAP3 is shown in 
Figure~\ref{10Pk3contours_3y}, where the difference in the improvements can be seen in more detail (compare to Fig.\ref{10Pk3contours}). 

\begin{table*}[t!]
\caption{\label{tab_results_10P_3y} Same as Table~\ref{tab_results_9P_3y} but for $w$CDM models. Compare with Table~\ref{tab_results_10P} for WMAP1.}
\begin{ruledtabular}
\begin{tabular}{l|lllll|ll}
  & W+P & W+B & W+P+B & W+P+B (no mix. cov.) & W+P+B+P$_L$ & W+P+P$_L$ & W+P$_L$ \\
\hline\hline
 & \multicolumn{5}{l}{$k_{\smax}^{MS}=0.2 \kMpc$} & \multicolumn{2}{l}{$k_{\smax}^{LRG}=0.2 \kMpc$}\\\hline\hline
$\Delta\omega_d$      &$0.0044$ &$0.0047$~($-6\%$) &$0.0035$~($26\%$)      &[~$0.0036$~($22\%$)~] &${\bf 0.0028}$~($57\%$) &$0.0032$~($37\%$)      &$0.0036$~($22\%$) \\\hline
$\Delta\omega_b$      &$0.00060$&$0.00061$~($-2\%$)&$0.00058$~($3\%$)      &[~$0.00059$~($2\%$)~] &$0.00057$~($5\%$)       &$0.00058$~($34\%$)     &$0.00059$~($2\%$) \\\hline
$\Delta\Omega_\Lambda$&$0.0104$ &$0.0092$~($13\%$) &$0.0073$~($42\%$)      &[~$0.0072$~($44\%$)~] &${\bf 0.0054}$~($93\%$) &${\bf 0.0062}$~($68\%$)&$0.0072$~($44\%$) \\\hline
$\Delta n_s$          &$0.0130$ &$0.0132$~($-1\%$) &$0.0126$~($3\%$)       &[~$0.0126$~($3\%$)~]  &$0.0121$~($7\%$)        &$0.0124$~($5\%$)       &$0.0128$~($2\%$)  \\\hline
$\Delta A_s$          &$0.037$  &$0.038$~($-3\%$)  &$0.035$~($6\%$)        &[~$0.036$~($3\%$)~]   &$0.035$~($6\%$)         &$0.036$~($3\%$)        &$0.037$~($0\%$)   \\\hline
$\Delta w$            &$0.060$  &$0.058$~($3\%$)   &$0.045$~($33\%$)       &[~$0.048$~($25\%$)~]  &${\bf 0.040}$~($50\%$)  &$0.046$~($30\%$)       &$0.051$~($18\%$)  \\\hline
$\Delta \tau$         &$0.025$  &$0.025$~($0\%$)   &$0.025$~($0\%$)        &[~$0.025$~($0\%$)~]   &$0.025$~($0\%$)         &$0.025$~($0\%$)        &$0.025$~($0\%$)   \\\hline
$\Delta b_1$          &$0.082$  &$0.101$~($-19\%$) &$0.058$~($41\%$)       &[~$0.069$~($19\%$)~]  &${\bf 0.050}$~($64\%$)  &$0.062$~($32\%$)       & -                \\\hline
$\Delta b_2$          & -       &$0.068$           &$0.051$                &[~$0.061$~]           &$0.047$                 & -                     & -                \\\hline
$\Delta b_1^{LRG}$    & -       & -                & -                     & -                    &$0.091$                 &$0.110$                &$0.121$           \\\hline\hline
$\Delta \sigma_8$     &$0.048$  &$0.050$~($-4\%$)  &$0.034$~($41\%$)       &[~$0.038$~($26\%$)~]  &${\bf 0.029}$~($66\%$)  &$0.036$~($33\%$)       &$0.041$~($17\%$)  \\\hline
$\Delta h$            &$0.0141$ &$0.0123$~($15\%$) &${\bf 0.0089}$~($58\%$)&[~$0.0096$~($47\%$)~] &${\bf 0.0074}$~($91\%$) &${\bf 0.0093}$~($52\%$)&$0.0107$~($32\%$) \\\hline
$\Delta \Omega_b$     &$0.00194$&$0.00168$~($15\%$)&$0.00134$~($45\%$)     &[~$0.00140$~($39\%$)~]&${\bf 0.00116}$~($67\%$)&$0.00134$~($45\%$)     &$0.00150$~($29\%$)\\\hline
\hline
 & \multicolumn{5}{l}{$k_{\smax}^{MS}=0.3 \kMpc$} & \multicolumn{2}{l}{$k_{\smax}^{LRG}=0.2 \kMpc$}\\\hline\hline
$\Delta\omega_d$      &$0.0042$ &$0.0038$~($11\%$) &$0.0030$~($40\%$)       &[~$0.0029$~($45\%$)~] &${\bf 0.0025}$~($68\%$) &$0.0032$~($31\%$)      & \\\hline
$\Delta\omega_b$      &$0.00059$&$0.00058$~($2\%$) &$0.00058$~($2\%$)       &[~$0.00056$~($5\%$)~] &$0.00056$~($5\%$)       &$0.00057$~($4\%$)      & \\\hline
$\Delta\Omega_\Lambda$&$0.0090$ &$0.0061$~($48\%$) &${\bf 0.0059}$~($53\%$) &[~$0.0052$~($73\%$)~] &${\bf 0.0048}$~($88\%$) &${\bf 0.0059}$~($53\%$)& \\\hline
$\Delta n_s$          &$0.0128$ &$0.0125$~($2\%$) &$0.0124$~($3\%$)        &[~$0.0120$~($7\%$)~]  &$0.0120$~($7\%$)        &$0.0123$~($41\%$)      & \\\hline
$\Delta A_s$          &$0.037$  &$0.037$~($0\%$)   &$0.034$~($9\%$)         &[~$0.035$~($6\%$)~]   &${\bf 0.033}$~($12\%$)  &$0.036$~($3\%$)        & \\\hline
$\Delta w$            &$0.057$  &$0.048$~($19\%$)  &$0.039$~($46\%$)        &[~$0.041$~($39\%$)~]  &${\bf 0.037}$~($54\%$)  &$0.046$~($24\%$)       & \\\hline
$\Delta \tau$         &$0.025$  &$0.025$~($0\%$)   &$0.025$~($0\%$)         &[~$0.025$~($0\%$)~]   &$0.025$~($0\%$)         &$0.025$~($0\%$)        & \\\hline
$\Delta b_1$          &$0.077$  &$0.085$~($-9\%$)  &${\bf 0.049}$~($57\%$)  &[~$0.057$~($35\%$)~]  &${\bf 0.043}$~($79\%$)  &$0.060$~($28\%$)       & \\\hline
$\Delta b_2$          & -       &$0.045$           &$0.039$                 &[~$0.044$~]           &$0.037$                 & -                     & \\\hline
$\Delta b_1^{LRG}$    & -       & -                & -                      & -                    &$0.079$                 &$0.108$                & \\\hline\hline
$\Delta \sigma_8$     &$0.046$  &$0.040$~($15\%$)  &${\bf 0.027}$~($70\%$)  &[~$0.029$~($59\%$)~]  &${\bf 0.024}$~($92\%$)  &$0.035$~($31\%$)       & \\\hline
$\Delta h$            &$0.0124$ &$0.0084$~($48\%$) &${\bf 0.0070}$~($77\%$) &[~$0.0069$~($80\%$)~] &${\bf 0.0063}$~($97\%$) &$0.0089$~($39\%$)      & \\\hline
$\Delta \Omega_b$     &$0.00173$&$0.00122$~($42\%$)&${\bf 0.00114}$~($52\%$)&[~$0.00113$~($53\%$)~]&${\bf 0.00106}$~($63\%$)&$0.00130$~($33\%$)     & \\\hline
\end{tabular}
\end{ruledtabular}
\end{table*}

\begin{figure*}[t]
\begin{center}
\includegraphics[width=0.98\textwidth~]{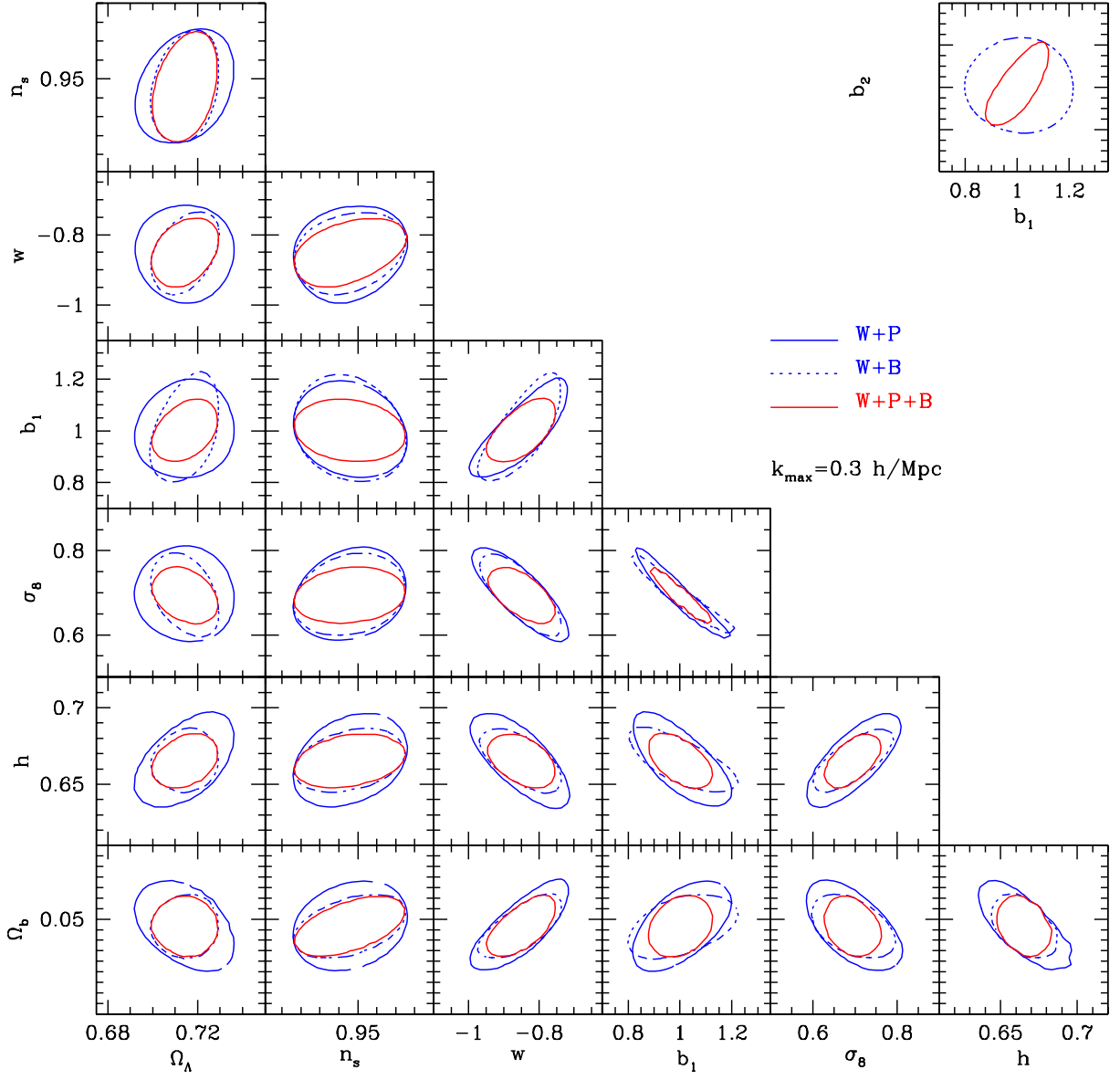}
\caption{$w$CDM models: marginalized $95\%$ contour plots for a selection of the cosmological parameters assuming $k_{\smax}^{MS}=0.3 \kMpc$. Compare with Fig.~\ref{10Pk3contours} for WMAP1.}\label{10Pk3contours_3y}
\end{center}
\end{figure*}

\end{document}